\documentclass[12pt]{article}
\usepackage{epsfig}
\usepackage{amssymb}
\usepackage{a4}
\usepackage{graphicx}
\evensidemargin \oddsidemargin

\newcommand{\be}{\begin{equation}}
\newcommand{\ee}{\end{equation}}
\newcommand{\ba}{\begin{array}}
\newcommand{\ea}{\end{array}}

\newcommand{\bea}{\begin{eqnarray}}
\newcommand{\eea}{\end{eqnarray}}

\newtheorem{prop}{Property}
\newtheorem{lemma}{Lemma}
\newtheorem{theorem}{Theorem}

\def\sq{\mbox{\rlap{$\sqcap$}$\sqcup$}}
\newenvironment{proof}[1]{\vspace{5pt}\noindent{\bf Proof #1}\hspace{6pt}}%
{\hfill\sq}
\newcommand{\bp}{\begin{proof}}
\newcommand{\ep}{\end{proof}\par\vspace{10pt}\noindent}
\def\b#1{{\mathbb #1}}

\def\nn{\nonumber  \\}

\begin{document}

\title{On kinks and other travelling-wave solutions of a modified
sine-Gordon equation}

\author{ Gaetano Fiore$^{1,2}$, \ Gabriele Guerriero$^1$, \ Alfonso Maio$^1$, \  Enrico Mazziotti$^1$ \\
~  \\
  $^1$Dip. di Matematica e Applicazioni, Universit\`a ``Federico II'',\\
 V. Claudio 21, 80125 Napoli\\
$^2$I.N.F.N., Sezione di Napoli, 
Complesso MSA, V. Cintia, 80126 Napoli\\
email: gaetano.fiore@unina.it}

\date{}

\maketitle
\abstract{We give an exhaustive, non-perturbative classification of
exact travelling-wave solutions of a perturbed sine-Gordon equation  (on the real
line or on the circle) which is used to describe the
Josephson   effect in the theory of superconductors and other
remarkable physical phenomena.  The perturbation  of the equation consists of
a constant forcing term and a linear  dissipative term. On the real line candidate
orbitally  stable solutions with bounded energy density are either the constant one,
or of kink (i.e. soliton) type, or  of array-of-kinks type, or of
``half-array-of-kinks'' type. While the first three have unperturbed
analogs, the last type is essentially new.   We also propose a convergent
method of successive approximations of the (anti)kink solution based on 
a careful application of the fixed point theorem.}



\section{Introduction}

\begin{figure}
\begin{center}\includegraphics[width=10cm]{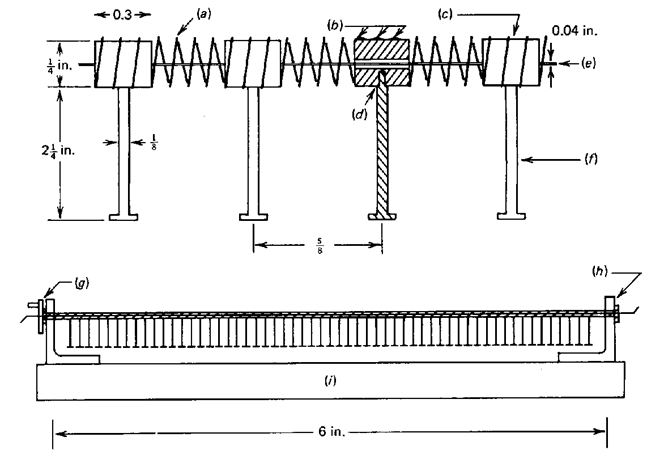}\hfill\includegraphics[width=6cm]{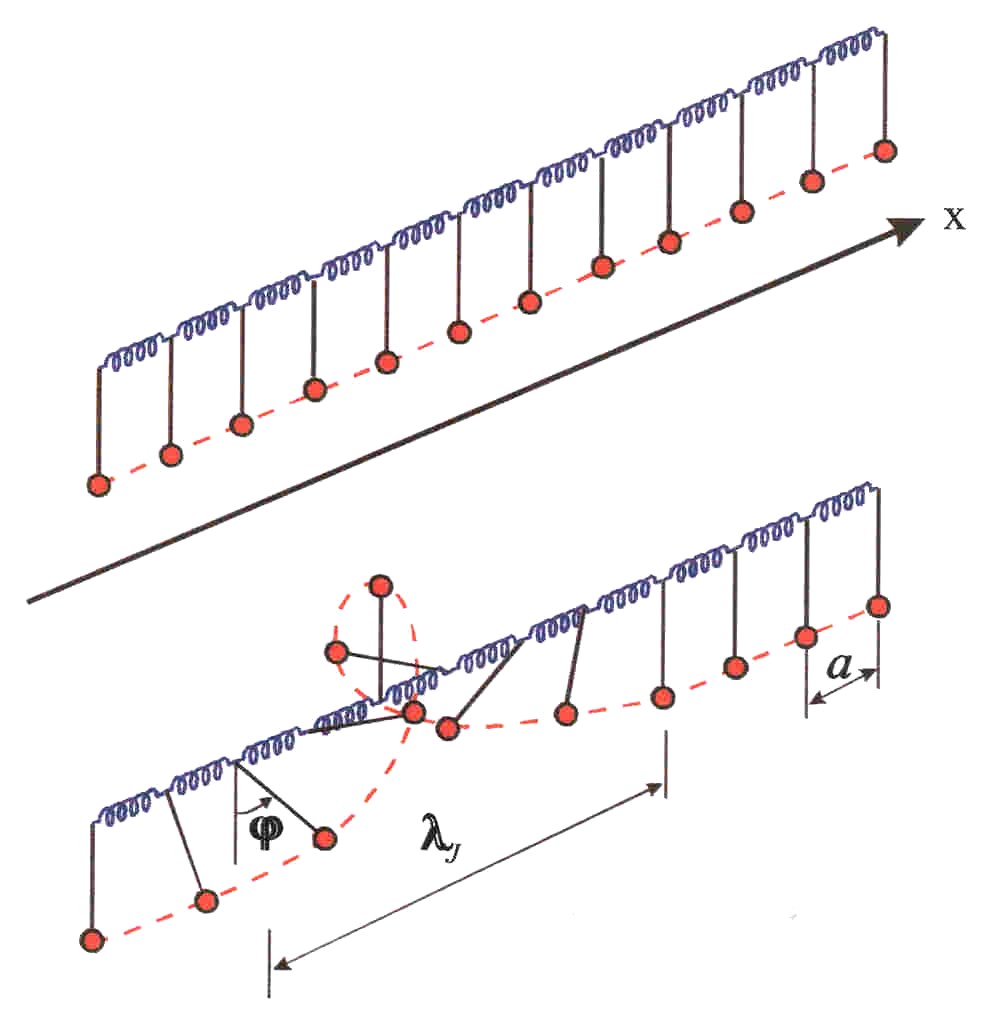}
\caption{Left: discretized mechanical model for the sine-Gordon equation, with (a) Spring, (b)
solder, (c) brass, (d) tap and thread, (e) wire, (f) nail, (g) and (h) ball
bearings, (i) base.
Right: stable static  (up) and kink (down) solutions.}  
\end{center}
\label{figura1}
\end{figure}

The purpose of this work is an exhaustive, non-perturbative analysis of
travelling-wave solutions of the ``perturbed'' sine-Gordon
equation
\be                               \label{equation}
\varphi_{tt}-\varphi_{xx}
    + \sin\varphi+\alpha \varphi_t+\gamma =0,
    \qquad x\in\b{R},
 \ee
for {\it all} constant $\alpha\!\ge \!0,\gamma\in \b{R}$.
This equation has been used to
describe with a good approximation a number of interesting physical phenomena,
notably Josephson effect in the theory of superconductors
\cite{Jos}, which is at the base \cite{BarPat82} of a large number of
advanced developments  both in fundamental research (e.g.
macroscopic effects of quantum physics, quantum computation) and in
applications to electronic devices (see e.g. Chapters 3-6 in
\cite{ChrScoSoe99}), or e.g. the  propagation of localized
 magnetohydrodynamic modes in plasma physics
\cite{Sco04}. The last two  terms are respectively a dissipative and a forcing
one; the  sine-Gordon equation (sGe) is obtained by setting them
equal to zero.
In the Josephson effect (for an introduction see e.g. Chapter 1 in \cite{BarPat82})
$\varphi(x,t)$ is the phase difference of the macroscopic quantum
wavefunctions describing the Bose-Einstein condensates of
Cooper pairs in two superconductors separated by a very thin, narrow
and  long dielectric (a socalled ``Josephson junction'').
The $\gamma$ term is the (external) ``bias current'',
providing energy to the system, whereas the dissipative term $\alpha
\varphi_t$ is due to Joule effect of the residual current across the junction
due to single electrons. We neglect the (tipically, rather small) 
surface losses of the superconductors, which are described by the more
complete equation
\be
-\varepsilon \varphi_{xxt}\!+\!\varphi_{tt}\!-\!\varphi_{xx}
    \!+\! \sin\varphi\!+\!\alpha \varphi_t\!+\!\gamma \!=\!0, \qquad\qquad 
\varepsilon\!>\!0.                                  \label{equation'}
\ee
Eq. (\ref{equation}) describes also the dynamics of
the continuum limit of a sequence \cite{Sco70} of pendula constrained
 to rotate around the same horizontal $x$-axis, subject to a constant torque $\gamma$, 
a viscous force  $-\alpha \varphi_t$ 
(due e.g. to their immersion in a viscous fluid) and coupled to each other
through a torque spring; $\varphi(x,t)$
is the deviation angle from the lower vertical position  at time $t$ of the
pendulum having position $x$ (see fig. \ref{figura1}).

It important to clarify: a) which solutions of the sGe
are deformed into solutions of (\ref{equation})
with the same qualitative features; b) whether (\ref{equation}) admits
also new kinds of solutions.
Candidate approximations to the former can be obtained
within the standard perturbative method (see e.g.
\cite{SatYaj74, KauNew76}
 and references therein) based on
modulations of the unperturbed solutions with slowly varying
parameters (typically velocity, space/time phases, etc.) and small radiation
components. In particular,
the Ansatz for a deformation of a travelling-wave solution
$\varphi_{(0)}(x,t)\!=\! g_{(0)}(x\!-\!vt)$ of the sGe  reads
\be
\varphi(x,t) =  g_{(0)}\Big(x\!-\!x_0(t)\!-\!\tilde v(t)t\Big) \!+\!
\gamma\varphi_{(1)}(x,t)\!+\!... \, ;                    \label{modulation}
\ee
$\gamma$ plays the role
of perturbation parameter, whereas the slowly
varying $x_0(t),\tilde v(t)$  and the perturbative ``radiative'' corrections
$ \gamma\varphi_{(1)}(x,t)+...$ have to be computed perturbatively in
terms    of  $\alpha \varphi_t+\gamma $.
If in particular $\varphi_{(0)}(x,t)$ is a (anti)kink [equivalently, (anti)soliton] solution,
one finds \cite{FogTrulBisKru76,McLSco77} also candidate approximate solutions
with {\it constant} velocity
\be                               \label{v_atinf}
\tilde v(t)\equiv v_{\infty}:=\pm[1+(4\alpha/\pi\gamma)^2]^{-\frac 12}
\ee
which are characterized by a power balance between the dissipative term
$\alpha \varphi_t$ and the external force term $\gamma$. 
Studying the convergence of the perturbative series would be difficult, not 
very illuminating for the existence and qualitative properties of the solutions for 
small $\gamma$, and certainly inadequate  for large $\gamma$.
Numerical resolutions \cite{Joh68,NakOnoNakSat74} of
(\ref{equation}) are  indicative,  but cannot provide solid,
exhaustive answers to the two questions above.

The travelling-wave Ansatz transforms (\ref{equation}) into a well-known \cite{Tri,Ame49,SanCon56,AndCha49}
Ordinary Differential Equation (ODE), whose phase space analysis in principle allows to
give a complete classification (and additional qualitative properties) of the 
travelling-wave solutions. However, up to our knowledge previous works 
\cite{Mag80,DerDoeAvanVis03} present an incomplete classification, 
in that they consider only solutions of type a) and they stick only to  part 
(near the origin) of the parameter space $(\alpha,\gamma)$.
The main purpose of this work is to provide (section \ref{PSG}) 
an exhaustive non-perturbative {\it classification of exact travelling-wave}
solutions of  (\ref{equation})  on the real line or on
the circle   {\it for all $\alpha\!\ge \!0,\gamma\in \b{R}$} and to propose
(section \ref{successapprox}) an improved
method of successive approximations converging to the (anti)kink
solutions, at least for sufficiently small $\gamma$. As we will see, the classification
includes also solutions of type b).
We will in particular concentrate on solutions of physical interest, namely solutions that have bounded energy density $h$ (and therefore also
bounded derivatives) and are
known, or are candidate, to be orbitally stable;
 in the sequel we shall denote them
as the {\it relevant solutions}. 
To make the paper essentially self-contained,  we give detailed
preliminaries in sections \ref{Preliminary},\ref{PSG} and in the appendix. 
If the velocity is  $\pm 1$ the ODE is
of first order and can be solved by quadrature,
otherwise it is the second order one
describing the motion along a line of a particle subject to a ``washboard''
potential and immersed in a linearly viscous fluid (or equivalently a single pendulum subject to
a constant torque and to a viscous force), and therefore the
problem is essentially reduced to studying this simpler mechanical analog.
Several useful monotonicity properties (subsection \ref{monoproperties})
allow in particular to identify   (Theorem 1 in section  \ref{PSG})
four families of relevant solutions:
three of them (the arrays of kinks for all values of $\gamma$,
the kinks and the constants only for
$\gamma\!<\! 1$) are deformations of analogous families  of
solutions of the sGe, whereas the fourth family
is without unperturbed analog: as each of its elements
interpolates between  a kink
and an array of kinks (see Fig. \ref{summaryfig2}),
we have named it a ``half-array of  kinks''.
None of the other solutions is relevant.
The families of perturbed kinks and arrays of kinks
depend on one free parameter less than the unperturbed
ones, as the phase velocity $v$ becomes a
function of the other parameters [for the kink coincides,
at lowest order in $\gamma$, with (\ref{v_atinf})].

Theorems of  existence and uniqueness for a class of 
dissipative equations including (\ref{equation'})  and the following boundary conditions 
have been proved: $\varphi\!\to\!0$  as $x\!\to\!\pm\infty$  \cite{DacRen92,DeaRen08};  
Dirichlet, Neumann or pseudoperiodic conditions
on a {\it  finite} space interval \cite{DeAFio13}. The boundedness and stability of
 solutions for this class  with  Dirichlet boundary conditions 
has been studied in \cite{DanFio05-13,Rio12}. 
The study of (\ref{equation}-\ref{equation'}) on the whole real axis is  more difficult;
at the end of section \ref{PSG} we  briefly recall some results 
\cite{Sco69,JonMarMilPla13,Mag80,DerDoeAvanVis03} on the stability
of travelling-wave solutions of (\ref{equation}) which can be found in the literature.
As we shall briefly explain,  a comprehensive stability analysis requires additional work.

\section{Preliminaries}
\label{Preliminary}

Space or time translations transform any solution $\varphi(x,t)$ into a new one
$\varphi(x\!+\!a,t\!+\!b)$; one thus obtains a two-parameter family of solutions  $[\varphi]$.
The pendula chain model described above allows a qualitative comprehension of the main
features of the solutions, e.g. of their instabilities.
The constant solutions of (\ref{equation})
are $\varphi^s(x,t)\equiv - \sin^{-1}\gamma\!+\!2\pi k$
and $\varphi^u(x,t)\equiv  \sin^{-1}\gamma\!+\!(2k\!+\!1)\pi $. The former
are stable, the latter unstable, as they yield respectively local minima and
maxima of the energy density
\be
h:=\frac {\varphi_t^2}2+
\frac {\varphi_x^2}2+\gamma\varphi-\cos\varphi.         \label{endensity}
\ee
They resp. correspond to configurations with all pendula hanging down or standing up.
Our {\it definition of a kink} $\varphi$
is:   $\varphi$ is a non-constant {\it stable travelling-wave
solution with 
all derivatives rapidly going to zero outside some localized region}. Then
mod. $2\pi$ it must be
\be                                     \label{asymcond}
\lim\limits_{x\to \!-\!\infty}\varphi(x,t)\!=-\!\sin^{-1}\!
\gamma,       \qquad
 \lim\limits_{x\to \!+\!\infty}\varphi(x,t)=-\!\sin^{-1}\!\gamma+\! 2n\pi
\qquad \ee
with $n\in\b{Z}$.
As we shall recall below, only $n=1$ (kink)  and
$n=-1$ (antikink) are possible
[whereas $n=0$ corresponds to the constant $\varphi_s$].
In the mentioned mechanical model the kink (resp. antikink) solution
describes a localized twisting  of the pendula chain by $2\pi$ around the axis anticlockwise
(resp. clockwise),
which moves with constant velocity.
The above condition yields an energy density $h$
(rapidly) going to two local minima as
$x\!\to\! \pm\infty$.
Although this makes the total Hamiltonian
$H\!:=\!\int_{-\infty}^{+\infty}h(x,t)dx$
divergent, the time-derivative is finite and non-positive:
\vskip-.3cm
$$
\qquad\qquad \qquad\qquad \qquad \dot H=-\int^{\infty}_{-\infty}\!\!\alpha\varphi_t^2dx
\le 0.
$$
[The negative sign at the right-hande side (rhs) shows the dissipative character of the 
$\alpha$-term in (\ref{equation})]. The effect of $\gamma\neq 0$
is to make the values of the energy potential density \ $\gamma\varphi\!-\!\cos\varphi$ \ at any two minima
different; this leaves room for a steady compensation of the energy dissipated by the $\alpha$-term
and the variation of the total potential energy due to the substantial variation of $\varphi(x,t)$ from
one minimum point to  the lower next in some spacial interval, and so may account for solutions not being
damped to constants  as $t\to\infty$.

\medskip
Without loss of generality we can assume $\gamma \ge 0$. If originally
this is not the case, one just needs to replace $\varphi\to -\varphi$.
If $\gamma> 1$ no solutions $\varphi$ having
finite limits and vanishing derivatives for $x\to \pm \infty$  can exist, in particular
no static solutions. If $\gamma=1$ the only static solution $\varphi$ having for
$x\to \pm \infty$  finite limits and vanishing derivatives is
$\varphi\equiv -\pi/2\mbox{ (mod }2\pi)$, which however is manifestly unstable.

\subsection{Reduction to ODE by the travelling-wave Ansatz}
\label{PDEODE}

We specify our travelling-wave Ansatz as follows:
\be
\ba{lll}
\xi\!:=\!\pm  x\!-\! t\qquad
\quad
& \varphi(x,t)\!=\!g(\xi)\!-\!\pi \:\:\:\quad &\mbox{ if }v\!=\!\pm 1, \\[8pt]
\xi\!:=-\mbox{sign}(v)\frac{x\!-\!vt}{\sqrt{v^2\!-\!1}}\qquad
\quad
& \varphi(x,t)\!= \!- g(\xi) \:\:\:\quad &\mbox{ if }v^2\!> \!1, \\[8pt]
\xi\!:=\!\mbox{sign}(v)
\frac{x\!-\!vt}{\sqrt{1\!-\!v^2}}
\qquad \quad & \varphi(x,t)\!=\!g(\xi)\!-\!\pi
\:\:\:\quad&\mbox{ if } 0\!<\!v^2\!<\! 1,\\[8pt]
\xi\!:=
\!x\qquad\quad & \varphi(x,t)\!=\!g(\xi)\!-\!\pi \:\:\:\quad &\mbox{ if
}v=0. \ea              \qquad                    \label{redef'}
\ee
If $v=\pm 1$, replacing the Ansatz
 in (\ref{equation}) one obtains the {\bf first order ODE},
\be
\alpha  g' = \gamma-\sin  g.  \label{equa}
\ee
One can explicitly solve this equation by quadrature \cite{Fio08}.
Already in \cite{DanDeaFio05} it has been argued that if $\gamma\!<\!1$
all solutions of (\ref{equa}) yield unstable solutions of (\ref{equation}), except the  constant one,
which yields the static constant solution $\varphi^s(x,t)\equiv-\sin^{-1}\gamma$. 
The same argument holds also if $\gamma=1$. If $\gamma\!>\!1$, by integrating
one finds
$$
\xi-\xi_0=\int^{\xi}_{\xi_0}d\xi'=
\alpha\int^{
g}_{ g_0} \frac{ds}{\gamma-\sin s};
$$
the denominator is positive for all $s\in\b{R}$, so that the solution $g$
is strictly monotonic and pseudoperiodic, i.e. the sum of a linear and a
periodic function, so that (for all $\xi,g_0$)
\be
g(\xi+\Xi)=  g( \xi)+2\pi,  \qquad\qquad\Xi:=  \int^{g_0+2\pi}_{g_0} \frac{ds}{u(s)}
\label{linper}
\ee
with $u(g)\!:=\!g'(g)\!=\!(\gamma\!-\!\sin g)/\alpha$. We shall
denote such solution as $\tilde g(\xi)$; this will yield (Theorem
\ref{ThmClassif}) a candidate stable solution $\check\varphi$ of (\ref{equation}),
representing an `array of (anti)kinks' travelling with velocity $\pm 1$
(such velocities are not possible in the sGe case).

\begin{figure}
\begin{center}\includegraphics[width=8cm]{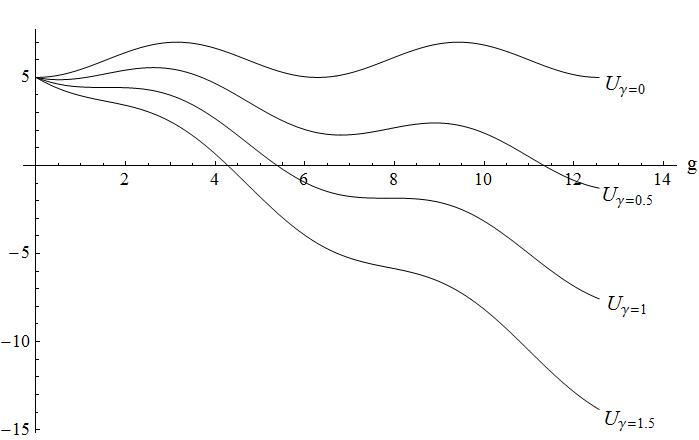}
\caption{The potential energy $U(g)=6\!-\!(\cos g\!+\!\gamma g)$ for  $\gamma=0$, $\gamma=.5$,
$\gamma=1$, $\gamma=1.5$.}  
\end{center}
\label{figura2}
\end{figure}

\bigskip
In the rest of the section we assume that
$v\neq \pm 1$.
Replacing in (\ref{equation}) we find in all three
remaining cases the {\bf second order} ODE
\be                               \label{equation"}
g''+\mu g'+U_g(g)=0,
    \qquad \xi\in\b{R},
\ee
which can be regarded as the 1-dimensional equation of
motion  w.r.t. the `time' $\xi$ of a particle  with unit mass,
position $g$,
subject to a `washboard' potential energy'
$U(g)$ and a viscous force with viscosity coefficient $\mu$ given by
\be                                       \label{defmu}
U(g):=-(\cos g+\gamma g),\qquad\qquad\mu:=\frac{\alpha}{\sqrt{|v^{-2}-1|}}.
\ee
Note that in equation (\ref{equation"}) $\alpha, v$ appear
only through their combination (\ref{defmu})$_2$, and that
 in the range $|v|\in[0,1[$ (resp. $|v|\in]1,\infty[$) $\mu(|v|)$ is
strictly increasing (resp. decreasing), and therefore
invertible.
In Fig. \ref{figura2} $U(g)$ is plotted  for four different
values of $\gamma$; it admits local minima (resp. maxima) only if $0\le\gamma
<1$, in  the points
$$
g_k^m:=\sin^{-1}\!\gamma\!+ \!2k\pi,\qquad\qquad \qquad(\mbox{resp. }
g_k^M:=-\!\sin^{-1}\!\gamma\!+\!(2k\!+\!1)\pi).
$$
As $\gamma\to 1$ the points $g_k^m,g_k^M$  approach
each other, and for $\gamma=1$ $g_k^m=g_k^M=(2k+1/2)\pi$  are
inflections points. For $\gamma>1$ no minima, maxima or inflections
exist, and $U_g<0$ everywhere. The ``total energy
of the particle''
$\mbox{\rm e}:=g'{}^2/2\!+\!U(g)$ is a non-increasing
function of $\xi$, as ${\rm e}'=-\mu g'{}^2$.

An exhaustive classification of the solutions of equation
(\ref{equation"}) for all values of $\mu,\gamma$ has been performed
 in several works, starting from \cite{Tri,Ame49} (see e.g.
\cite{SanCon56} or \cite{AndCha49} for comprehensive presentations).
The equation is equivalent to the autonomous first order system \be
\ba{l}
u'=-\mu u-\sin g+\gamma,\\
g'=u. \ea          \label{eqsystem}
\ee
Since the rhs's are
functions of $g,u$ with bounded continuous derivatives, by the
Peano-Picard theorem on the extension of the integrals all solutions
are defined on all $-\infty<\xi<\infty$  (global existence), and 
the trajectories (paths) in the phase space $(g,u)$ do not intersect
(uniqueness). Each is uniquely identified by any of its points $(g_0,u_0)$.
The paths may have finite endpoints
(limits as $\xi\to\pm \infty$) only at
singular points [i.e. where the rhs's(\ref{eqsystem}) vanish].
These exist only for $\gamma\le 1$, lie all on the axis $u=0$,  and are
\be
\ba{ll}
\mbox{saddles } A_k=(g_k^M,0),\qquad\quad
\mbox{nodes, foci or centers } B_k=(g_k^m,0),\qquad\qquad&\gamma<1,\\[8pt]
\mbox{saddle-nodes } C_k=\big((2k+1/2)\pi,0\big) &\gamma=1.\ea
\ee
Their characteristic equations can be summarized as
\be 
\lambda^2+\mu\lambda \mp\sqrt{1\!-\!\gamma^2}=0; 
\ee 
the upper, lower sign resp. refer to the
$A_k$, $B_k$ for $\gamma\!<\!1$; $\gamma\!=\!1$ gives the solutions at $C_k$:
\begin{enumerate}

\item The solutions $\lambda_1,\lambda_2$ for $A_k$ are real of opposite sign,
and  $A_k$ is a saddle point.

\item The solutions $\lambda_1,\lambda_2$ for $B_k$ are:
\begin{itemize}

\item Both real and negative 
if $\mu\ge 2(1\!-\!\gamma^2)^{1/4}$.  $B_k$ is a node, and there are
an infinite number of paths going to $B_k$ as $\xi\!\to\!\infty$ with the same
tangent. These represent overdamped motions of the `particle' towards $g_k^m$.

\item Complex conjugates (but not purely imaginary) if $0<\mu <
2(1\!-\!\gamma^2)^{1/4}$.  $B_k$ is a focus, and there are an
infinite number of paths  going to $B_k$  along a spiral as $\xi\!\to\!\infty$. These
represent damped oscillations  of the `particle' about $g_k^m$.

\item Opposite imaginary if $\mu=0$. $B_k$ is a center, and
there exist closed paths (cycles) around it. These represent periodic oscillations  of
the `particle' about $g_k^m$.

\end{itemize}

\item If $\gamma\!=\!1$ then $\lambda_1\!=\!0$, $\lambda_2\!=\!-\mu$. 
If $\mu\!>\!0$ $C_k$ is a saddle-node: there are two 
separatrices in the half-plane $g\!>\!(2k+1/2)\pi$ (one going  leftwards to  $C_k$,
the other leaving from $C_k$ rightwards) and infinitely many in the half-plane
$g\!<\!(2k+1/2)\pi$ going rightwards to $C_k$ (these again represent overdamped 
motions of the `particle'). 
\end{enumerate}
The solutions are continuous functions of the
parameters $\mu,\gamma$ and of $(g_0,u_0)$ (away from singular
points), uniformly in every compact subset.
In subsection \ref{monoproperties} we recall that the latter dependences are also monotonic.

\medskip

To analyze the qualitative features,
the monotonicity properties and the asymptotic behaviour of the paths
near the endpoints it is
useful to eliminate the `time' $\xi$ and adopt as an independent variable the
`position' $g$,  as in the unperturbed case.
The path of any solution $g(\xi)$ of (\ref{equation"}) is cut into
pieces by the axis $u=0$. Let $Y\equiv]\xi_-,\xi_+[\subseteq \b{R}$
be the 'time' interval corresponding to a piece,
$$
\epsilon:=\mbox{sign}\big(u(\xi)\big),\qquad \qquad \xi\in Y
$$
be its sign
and let $G\equiv]g_-,g_+[:=g(Y)$. In $Y$ the function $g(\xi)$ can
be inverted to give a function $\xi:g\in G\to \xi(g)\in Y$. So one
can express  the `velocity' $u$ and the `kinetic energy' $z:=u^2/2$
of the `particle' as functions of its `position' $g$.  By derivation
we find that $g''(\xi)=u_g\big(g(\xi)\big) g'(\xi)$ and the second
order problem (\ref{equation"}) with initial condition
$\big(g(\xi_0),u(\xi_0)\big)\!=\!(g_0,u_0)$
in $Y$ is equivalent to two {\bf first order problems}: the first is
\be
z_g(g)+\epsilon\mu \sqrt{z(g)}+\sin
g-\gamma=uu_g(g)+\mu u(g)+\sin
g-\gamma=0, \qquad\qquad u(g_0)\!=\!u_0\qquad \label{1stord}
\ee
(note that this is invariant
under the replacement  $g\to g+2\pi$), which has to be solved first, and
yields a solution $u=u(g;g_0,u_0;\mu,\gamma)$
continuous in all arguments (away from singular points); the second is
\be
g'(\xi)=u\big(g(\xi)\big), \qquad\qquad g(\xi_0)=g_0,
\label{1storder'}
\ee
is integrated out by quadrature
\be
\xi-\xi_0=\int^{\xi}_{\xi_0}d\xi'=\int^{ g}_{ g_0}
\frac{ds}{u(s)}=\epsilon\int^g_{ g_0}
\frac{ds}{\sqrt{2z(s)}} \label{quadrature}
\ee
and implicitly yields a
solution $g=g(\xi;g_0,u_0;\mu,\gamma)$ in $Y$. If $Y$ is not the
whole $\b{R}$, the final step is the patching of solutions in
adjacent intervals $Y$.

Choosing in (\ref{quadrature}) $g$ as  $g_{\pm}$  one  obtains
$\xi_{\pm}$.
If $z(g)$ vanishes as $\eta^a\!:=\!|g_{\pm}\!-\!g|^{a}$ with $a\ge 2$ as $g\uparrow
g_+$ or $g\downarrow g_-$, then $\xi_+=\infty$
or $\xi_-\!=\!-\!\infty$.  The
behaviour of $u(g),z(g)$ near $g_{\pm}$ can be determined immediately solving
(\ref{1stord}) at leading order in a left (resp. right) neighbourhood of $g_+$
(resp. $g_-$).
In particular, if $\gamma\!<\!1$ and $g_{\pm}=g_k^M$ (a maximum point of
$U$) then the equation obtained by replacing the power law Ansatz
$u(g)\!=\!\eta^{a/2}u_{\pm}\!+\!o(\eta^{a/2})$ in (\ref{1stord}) is solved by
\be
\ba{ll}  u(g)\approx (g_+\!-\!g)u_{+\epsilon}\qquad
&\mbox{as }g\uparrow g_+,
\\[8pt] u(g)\approx (g\!-\!g_-)u_{-\epsilon}\qquad &\mbox{as }g\downarrow
g_-, \ea\qquad \label{nearA_k}
\ee
where
for $\epsilon,\epsilon'\in\{+,-\}$ $u_{\epsilon'\epsilon}$ is
defined by
$$
u_{\epsilon'\epsilon}:=\frac 12\left(
\epsilon'\mu+\epsilon\sqrt{\mu^2+4\sqrt{1\!-\!\gamma^2}}\right).
$$
Formula (\ref{nearA_k}) gives the leading behaviour of the four separatrices
having an end on $A_k$.

\bigskip
Problem (\ref{1stord}) is also
equivalent to  the {\bf Volterra-type  integral equation}
\be
z(g)=z_0\!+\!U(g_0)\!-\!U(g)\!-\epsilon\!\int\limits_{g_0}^{g}ds\,\mu\sqrt{2z(s)}
\label{volterra}
\ee
where $z_0:=u_0^2/2$. When $\mu=0$ (no dissipation) this
gives the solutions explicitly and amounts  to the
conservation of the `total energy' \ $\mbox{\rm e}(g)=z(g)\!+\!U(g)$ \ of the  `particle'.


\subsection{Monotonicity properties}
\label{monoproperties}

In agreement with the physical intuition, {\bf the solutions  of (\ref{1stord})
and the extremes of $G$ depend on the
parameters $\mu,z_0,\gamma$  monotonically} (see e.g.  \cite{Tri,Ame49}). 
For completeness, in the Appendix we recall the proof of
the following monotonicity properties.

\begin{prop}
As functions of $z_0$: $z=u^2/2$ is
strictly increasing;
$g_+$  is increasing and $g_-$
decreasing (strictly as long as they have not reached the values $\pm \infty$).
\label{mono2}
\end{prop}

\begin{prop}
As a function of both $\mu,-\epsilon\gamma$ the solution
$u(g;g_0,u_0;\mu,\gamma)$ is  strictly decreasing (resp.
 strictly increasing) for  $g\in]g_0,g_+[$ (resp. $g\in]g_-,g_0[$).
Correspondingly, the solution $g(\xi;g_0,u_0;\mu,\gamma)$ is
strictly decreasing as a function of both $\epsilon\mu,-\gamma$, and
so is either extreme  $g_{\pm}$ (strictly as long as it has not
reached values   $\pm \infty$).
 \label{mono1}

\end{prop}

\noindent
{\bf Remark.}
 In general $g_{\pm}$ will be discontinuous functions of
$\mu,z_0,\gamma$ at  $g_{\pm}=g_k^M$.

\medskip
\noindent
Whenever the domain $G$ of the solution
$z(g)$ contains  a whole interval $]g,g\!+\!2\pi[$ we define
\be
I(z,g):=\int\limits^{g\!+\!2\pi}_{g}ds
\sqrt{2z(s)}\label{defint}
\ee
Given any $g_0\in \bar G$,  let
$g_k:=g_0\!+\!2\pi k$, $K:=\{k\!\in\!\b{Z}~|~ g_k\!\in \!\bar G\}$ and
$I_k:=I(z,g_k)$ if $k,k\!+\!1\in K$.

\begin{prop}
If $\epsilon\!=\!-$ the sequences $\{z(g_k)\},\{I_k\}$ are strictly increasing
and diverging as $k\to \infty$, with $K$
bounded from below. If $\epsilon\!=\!+$
the sequences $\{z(g_k)\},\{I_k\}$ are: either constant, with
$K=\b{Z}$; or  strictly increasing and converging as $k\to \infty$, with $K$
bounded from below; or strictly decreasing,
diverging as $k\to -\infty$, and either converging as $k\to \infty$,
or with $K$ upper bounded.  Moreover,
\be
\label{bound} z(g_{k\!+\!1})-z(g_k)=2\pi\gamma-\epsilon\mu I_k.
\ee
\label{monoseq}
\end{prop}

\section{Classification of the 
 solutions}
\label{PSG}

\subsubsection*{Short reminder about the sine-Gordon equation}
\label{rsGe}

If \ $\gamma\!=\!\alpha\!=\!\mu\!=\!0$ \ (sGe) the `total energy of the particle'
$\mbox{\rm e}$ is conserved and its value  (together with the free
parameter $v$) parametrizes
 different kinds of solutions of (\ref{equation"}). Plotting $U(g)$ (fig.
\ref{summaryfig1} left) we get an immediate
 qualitative understanding of them.
 They all have bounded $z(g)\!=\!\mbox{\rm e}\!-\!U(g)$, and  therefore 
bounded $g'$. This implies that also
the  corresponding $\varphi_x,\varphi_t,h$ are bounded functions of $x,t$.
Only solutions  corresponding to  $\mbox{\rm e}\!\ge 1$ and any $v \!\in ]\!-\!1,1[$
are spectrally stable \cite{Sco69,BarEspMagSco71,
JonMarMilPla13}.
If $\mbox{\rm e}\!=\! 1$ a path either degenerates to a saddle point [e.g. $\big(g(\xi),u(\xi)\big)\!\equiv\! \big(g_0^M,0\big)\!=\!A_0$: the `particle' stays at $A_0$]
or is heteroclinic  (i.e. starts and ends
at  two neighbouring saddle points, e.g. $A_0,A_1$:
the `particle', confined in the interval $g_0^M\!<\!g\!<\!g_1^M$,
starts at `time' $\xi=-\infty$ from $A_0$
and reaches $A_1$ at $\xi=\infty$, or viceversa, see fig. \ref{summaryfig1} left). 
Replacing the result in (\ref{redef'}), mod. $2\pi$ they translate into
unstable solutions of the  sGe if $v^2\!>\!1$
(in the model of fig. \ref{figura1}
all pendula of the chain stand upwards outside a small region), and the 
celebrated families of (spectrally  stable) solutions
\be
\hat \varphi^{\pm}_{(0)}(x,t;v) =  4 \tan^{-1}\left\{\exp\left[\pm
\frac{x-vt}{\sqrt{1\!-\!v^2}}\right]\right\}\label{kink}
\ee
if $v^2\!<\!1$. In the model of fig. \ref{figura1}
all pendula of the chain hang downwards outside a small region, and
 within the latter they twist  around the $x$-axis $n=\pm 1$ times, i.e. once anti-clockwise or
clockwise, depending on the sign. $\hat\varphi_{(0)}^+(x,t;v)$
is the family of kink solutions,
$\hat\varphi_{(0)}^-(x,t;v)$ the family of antikink
solutions, parametrized by the  velocity $v$, which can take any value
in $]\!-\!1,1[$.

Similarly, unbounded orbits ($\mbox{\rm e}\!>\! 1$) correspond to arrays of 
kinks or antikinks if $v^2\!<\!1$. 
The corresponding solutions $\check g_{(0)}(\pm \xi;\mbox{\rm e})$ are pseudoperiodic, in the sense (\ref{linper}):
the `particle' travels towards the right from $g_-=-\infty$  to $g_+=\infty$
(or viceversa) and its `kinetic energy' $\check z(g)$ is $2\pi$-periodic (see fig.
\ref{summaryfig1} left), in particular
takes the same value $z_M$ at all points $g_k^M$,
\be
 \check z(g_k^M)=z_M \qquad\qquad\forall
k\in\b{Z}. \label{periodicity}
\ee
Again, the corresponding solutions of the sGe are
\cite{Sco69,BarEspMagSco71}
 unstable if $v^2\!> \!1$ and stable
if $v^2\!<\! 1$
 (`most' pendula  point resp. upwards and downwards   in the pendula chain model).
 The stable solutions
$\check\varphi_{(0)}^{\pm}(x,t)=\check g_{(0)}(\pm \xi;\mbox{\rm e})$ respectively describe
two-parameter families of  evenly spaced ``arrays of kinks and antikinks'',
the two parameters being the velocity $v$ hidden in (\ref{redef'}), which can take any value
in $]\!-\!1,1[$, and one of the variables $ \check{\mbox{\rm e}},z_M,\Xi_{(0)}$;
$\Xi_{(0)}$ is the 'time lapse' [computed by (\ref{linper})$_2$] 
 the 'particle' needs to travel a distance $2\pi$. 

Clearly there is a heteroclininc bifurcation at $ \mbox{\rm e}\!=\!1$,
or equivalently $ z_M\!=\!0$, or $\Xi_{(0)}\!=\!\infty$.

On the contrary, solutions corresponding to 
cycles around centers $B_k$ ($\mbox{\rm e} \!\in ]\!-\!1,1[$),
 or with $v^2\!>\!1$, are unstable  \cite{Sco69,BarEspMagSco71,
JonMarMilPla13}.



\subsubsection*{The perturbed  sine-Gordon equation}

If not all $\gamma,\alpha,\mu$ vanish (perturbed sine-Gordon) there are \cite{DanDeaFio05} solutions
$g(\xi)$ with  $g'$  diverging as $\xi$ goes to infinity\footnote{For instance, by Prop. \ref{monoseq}
if $\epsilon\!=\!-$ and $z(g)$ is defined at least in an interval of
length $2\pi$ then $g_+=\infty$, $z(g)$ diverges as $g\to \infty$,
$g'(\xi),\varphi_x,\varphi_t$ diverge as $\xi\to -\infty$.};  the corresponding
solutions $\varphi$ have  $\varphi_x,\varphi_t,h$  diverging at space and time
infinity, hence are not relevant.
In Ref. \cite{DanDeaFio05} we have analyzed all the possibilities for $\gamma\!<\!1$
 and shown (Prop. 1) that {\it relevant} (in the sense of
the introduction) solutions  $\varphi$, {\it  if they exist, can be} only of
 four types, all with $v^2\!\le\! 1$ and $\epsilon\!:=\!\mbox{sign} (g')\!\ge\! 0$; out of
them three are deformations of travelling-wave solutions  of the sGe.
Here we  show (Theorem \ref{ThmClassif}) that all four
types {\it  actually exist}, extending our analysis to {\it  all} the parameter space $(\alpha,\gamma)$.

We assume  (without loss of generality) $\gamma\!>\!0$, and for $\mu\!\in\![0,\infty]$ we set
\be
\check v(\mu):= \frac{\mu}{\sqrt{\alpha^2+ \mu^2}}\le 1,
\qquad\qquad\qquad
\xi^{\pm}(\mu):=\frac{\pm x\!-\!\check v(\mu)t}{\sqrt{1\!-\!v^2(\mu)}},
\qquad\mbox{if }
\check v(\mu)\!<\!1.
\label{arrvel}
\ee
[by definition, $\check v(\infty)\!=\!1$].

\medskip
For $\gamma\!\le\!1$, replacing in (\ref{redef'}) the constant solutions \  $g(\xi)\equiv A_k$, \ $g(\xi)\equiv B_k,C_k$, \
we resp. obtain the {\it stable} solution $\varphi^s$  and the {\it unstable} one $\varphi^u$
already given in section \ref{Preliminary}.

\begin{figure}
\begin{center}
\includegraphics[width=8.5cm]{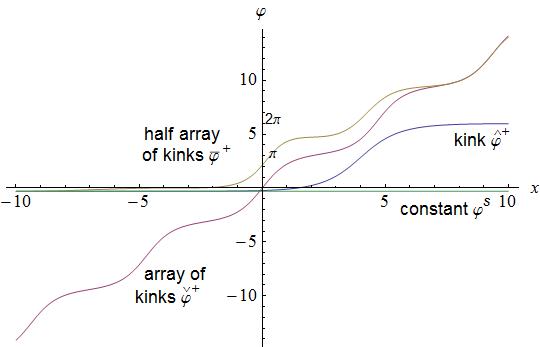}\hfill\includegraphics[width=8.5cm]{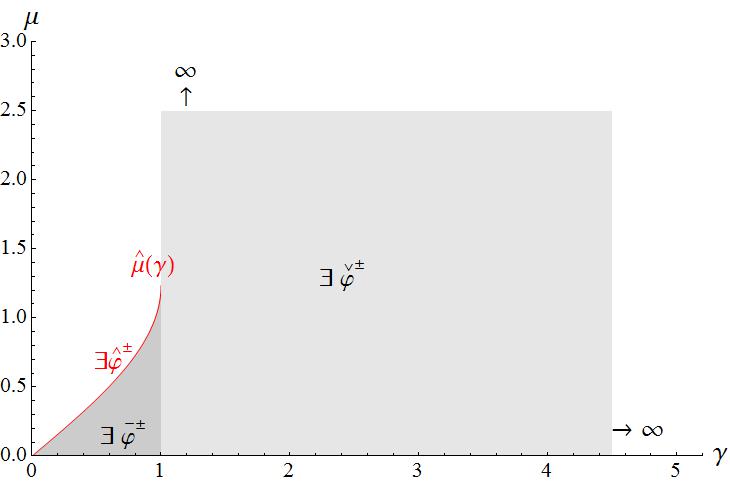}
\caption{Left: qualitative behaviour at $t\!=\!0$ of the following solutions of (\ref{equation}) for $\gamma\!=\!0.3$:
stable constant $\varphi^s$ (green), kink  $\hat \varphi^+(x\!-\!4,0)$  (blue), array of kinks  $\check \varphi^+(x,0)$ (violet)  and 
half-array of kinks  $\bar \varphi^+(x,0)$ (brown). 
Right: regions of the parameter space
where such solutions exist ($\varphi^s$ exists in the region $\gamma\!<\!1$).}
\label{summaryfig2}
\end{center}
\end{figure}

For a given $\mu\!>\!0$, a unique (up to a shift of $\xi$) pseudoperiodic path 
$\check p(\xi)\!=\!\big(\check g(\xi),\check u(\xi)\big)$ 
exists for sufficiently large $\gamma$; this attracts exponentially fast all 
other paths $p(\xi)\!=\!\big( g(\xi), u(\xi)\big)$ 
having $g_+\!=\!\infty$. 
In fact, since the two graphs  $z(g),\check z(g)$ do not intersect, $w(g)\!:=\!z(g)\!-\!\check z(g)$
 is either positive- or negative-definite. By (\ref{1stord})  it fulfills
\be
w_g=-\mu\left[\sqrt{2(\check z\!+\!w)}\!-\!\sqrt{2\check z}\right]
=-2\mu\frac{w}{\sqrt{2(\check z\!+\!w)}\!+\!\sqrt{2\check z}},                 \label{wg}
\ee
implying
$$
\frac d{dg}\ln |w|=-\frac{2\mu}{\sqrt{2(\check z\!+\!w)}\!+\!\sqrt{2\check z}}
\le -\frac{\mu}{\sqrt{2(\check z^M\!+\!|w(g_0)|)}}
$$
(we have denoted as $\check z^M$ the maximum of $\check z$ and as $g_0$
the initial $g$-point): $|w(g)|$ is strictly decreasing. By integration we find for $g\ge g_0$
\be
|w(g)|\le |w(g_0)|
e^{-C(g-g_0)}, \qquad\qquad
C:=\frac{\mu}{\sqrt{2(\check z^M\!+\!|w(g_0)|)}},
\ee
namely $|w(g)|\!\to\! 0$ exponentially fast as $g\!\to\!\infty$, as claimed. As $\gamma$ is
decreased, such an attracting path  disappears, becoming: a sequence of heteroclinic
paths connecting each $A_k$ with $A_{k+1}$, if $0 \!<\! \mu \!<\! \mu^*$;  
a saddle-node infinite-period bifurcation, if $ \mu \!>\! \mu^*$, with a special constant
$\mu^*$. Let $\hat\gamma(\mu)$ denote the bifurcation curve as a function of $\mu$. 
To our knowledge this curve has been first studied by Urabe \cite{Ura54},
who found $\mu^* \!\simeq\! 1.193$;
see e.g. also \cite{Str94,Lev78} and references therein for an updated report including more recent results.
$\hat\gamma(\mu)$  is a continuous function such that
$\hat\gamma(\mu) \!\simeq\! 4\mu/\pi$ as $\mu\!\sim\!0$ and $ \hat\gamma(\mu) \!=\!1$ for $\mu\!\ge\!\mu^*$.
In $[0,\mu^*]$ it is strictly increasing, hence invertible: we shall denote as $\hat\mu(\gamma)$  the inverse function.
This fulfills the bounds (\ref{hatmubounds}) and can be determined with
arbitrary accuracy for small $\mu$ also by the method described in  Theorem  \ref{born}. 
The above curves   play a crucial role in singling out different regions in the parameter space $(\gamma,\mu)$,
as depicted in fig. \ref{summaryfig2}-b. 

\begin{figure}
\begin{center}
\includegraphics[width=8cm]{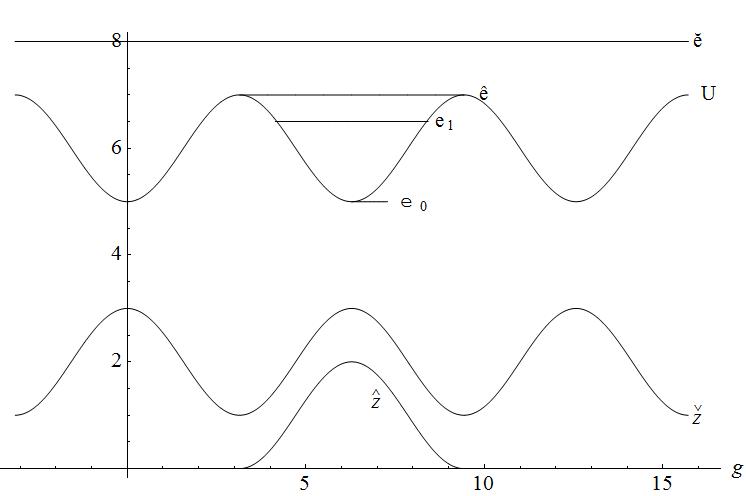}
\includegraphics[width=8cm]{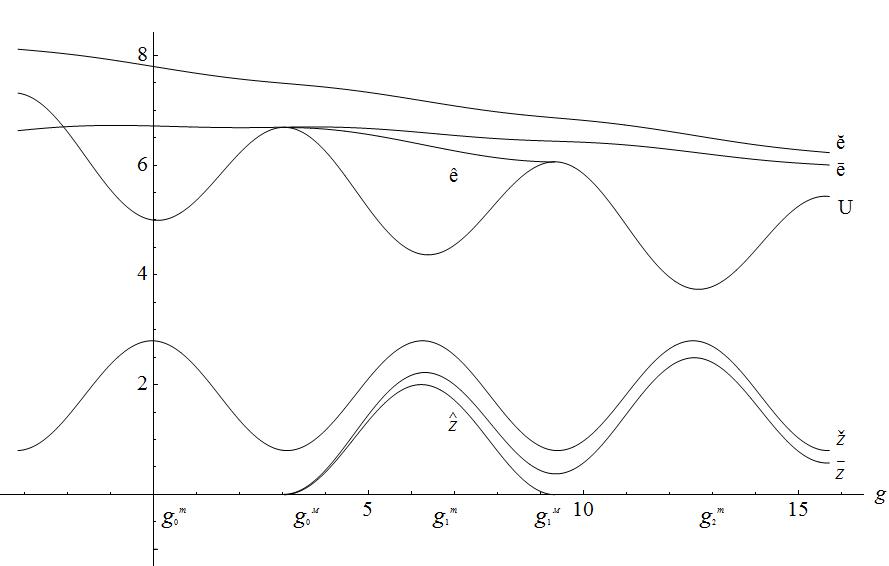}
\caption{The potential energy $U(g)=6 -(\cos g\!+\!\gamma g)$
for  $\gamma=0$ (left) and $\gamma=.1$ (right).
Correspondingly, the `kinetic energies' and  the `total
energies': 1)  $\hat z,\hat {\rm e}$  associated to the kink,
$\mu=\hat\mu(\gamma)$; 2) $\check z,\check {\rm e}$ associated to an
array of kinks, 
$\mu\!<\!\hat\mu(\gamma)$; 3) (only at the right) $\bar z,\bar {\rm e}$  associated to a half-array of
kinks solutions, $\mu\!<\!\hat\mu(\gamma)$.
}
\label{summaryfig1}
\end{center} \end{figure}

Fixed $\mu$, only for $\gamma \!>\!\hat\gamma(\mu)$ 
(light and dark grey areas in fig.  \ref{summaryfig2}-b) the attracting pseudoperiodic path $\check p(\xi)$ exists,
and $\epsilon\!=\!+$: the `particle' travels rightwards from $g_-=-\infty$  to $g_+=\infty$,
and its `kinetic energy' $\check z(g)$ not only fulfills  (\ref{periodicity}), but is $2\pi$-periodic
[see Fig. \ref{summaryfig1} right, where also $\check{\mbox{\rm e}}(g)$ is plotted].
By (\ref{bound}) this implies 
\be
\check\mu \,I\big(\check
z,g\big)=2\pi\gamma; \label{EnBalance}
\ee
the left-hand side (lhs) is
independent of $g$ (and can be called simply $\check I$). This
equality amounts to an {\bf energy balance condition}: `the energy dissipated by the viscous force 
equals the potential energy gap after a $2\pi$ displacement of the particle'.
For $g$ fixed, $\check z$, $\check I$  are strictly increasing, continuous
functions of $z_M$ by Proerties. \ref{mono2}, \ref{mono1}, whereas
$\check\mu$ and $\Xi$ are strictly decreasing and continuous
respectively by (\ref{EnBalance}) and (\ref{linper})$_2$. All
these functions are therefore invertible, and one can adopt any
of the four parameters $z_M,\check I,\mu,\Xi$ (in the
appropriate range) as the independent one, beside $\gamma$.
For $|v|\!<\!1$  one can adopt also $|v|$ as the independent
parameter, as the function $\mu(|v|)$
defined in (\ref{defmu})$_2$ is strictly monotonic.
As $\mu\!\to\!\infty$, or equivalently $|v|\!\to\!1$,  $\check g(\xi)$ goes to
the pseudoperiodic solution $\tilde g(\xi)$ of (\ref{equa}).
Replacing  $\check g(\xi)$ [or $\tilde g(\xi)$] in (\ref{redef'})
one finds one-parameter families of {\bf evenly spaced ``arrays of
kinks''} and of {\bf evenly spaced ``arrays of antikinks''},
as described in Theorem \ref{ThmClassif}; as a
parameter one can choose $z_M,\check I,\check\mu,\Xi$, or $|v|$.

For  $\mu\!<\!\mu^*$ and $\hat\gamma(\mu)\!<\! \gamma \!<\! 1$  
(dark grey region in fig.  \ref{summaryfig2}-b), the
pseudoperiodic attracting path $\check p(\xi)$ coexists with the sequence of alternating $B_k,A_k$.
For all $k\!\in\!\mathbb{Z}$, the saddle connection 
$\bar p_k(\xi)=\big(\bar g_k(\xi),\bar u_k(\xi)\big)$ that leaves from $A_k$
is attracted by $\check p(\xi)$ exponentially fast  (hence again $\epsilon\!=\!+$):
the `particle leaves at time $\xi=-\infty$ from  $g_k^M$ and its trajectory approaches more and more
$\check g(\xi)$ as $\xi\!\to\!\infty$'. 
Replacing  $\bar g_k(\xi)$ in (\ref{redef'})
one finds new one-parameter families of solutions,  the
 {\bf evenly spaced ``half-arrays of
kinks'' or ``antikinks''}, as described in Theorem \ref{ThmClassif}.

The part of the graph $\big(\hat\gamma(\mu),\mu\big)$ with $\mu\!\in\!]0,\mu^*]$ (the red curve in  
fig.  \ref{summaryfig2}-b) is a heteroclininc bifurcation:  fixed $\mu$, 
as $\gamma \!\downarrow\!\hat\gamma(\mu)$ the pseudoperiodic path approaches
the saddles, squeezing down - for all $k\!\in\!\mathbb{Z}$ - the saddle connection  $\bar p_k(\xi)$, 
and  for $\gamma \!=\!\hat\gamma(\mu)$  (i.e. on the red curve  in fig.  \ref{summaryfig2}-b) 
both $\check p,\bar p_k$ merge into a  heteroclinic
path $\hat p_k(\xi)=\big(\hat g_k(\xi),\hat u_k(\xi)\big)$ leaving from  $A_k$ and ending on
$A_{k+1}$ (hence again $\epsilon\!=\!+$):
the `particle, confined in the interval $g_k^M\!<\!g\!<\!g_{k+1}^M$,
leaves at time $\xi=-\infty$ from  $g_k^M$ and reaches  $g_{k+1}^M$ at time $\xi=\infty$'. 
The corresponding  `kinetic energy' $\hat z(g)$ is defined in the same interval
[see fig. \ref{summaryfig1} right, where also the corresponding
$\hat{\mbox{\rm e}}(g)$ is plotted]  and fulfills the boundary conditions
$\lim\limits_{g\downarrow g_k^M } \hat z(g) =0$, $\lim\limits_{g\uparrow g_{k+1}^M } \hat z(g) =0$. 
By (\ref{bound})  this implies
\be
\hat\mu \,I\big(\hat z,g_k^M\big)=2\pi\gamma, \label{EnBalances}
\ee
which is again the `{\bf energy balance condition} between the
energy dissipated by the viscous force and the potential energy gap
after a $2\pi$ displacement of the particle'.
Replacing $\hat g(\xi;\gamma)$  in (\ref{redef'})
one finds the {\bf perturbed (anti)kink solutions}, as
described in Theorem \ref{ThmClassif}.
The (anti)kink solution is also recovered from the array of (anti)kinks in the $z_M\!\to\! 0$ limit.
 By inversion of (\ref{defmu})$_2$
the velocity $v$ will be no more a free parameter, but the function $v=\pm\check v\big(\hat\mu(\gamma)\big)$
(with range $]\!-\!1,1\![$) of $\gamma,\alpha$ and  the helicity
$\pm$ of the (anti)kink solution $\hat\varphi^{\pm}$.

\smallskip
The  part of the graph $\big(\hat\gamma(\mu),\mu\big)$ with $\mu\!\in\!]\mu^*,\infty[$, i.e.
$\big(1,]\mu^*,\infty[\big)$ (the line separating the white from the light grey region in  
fig.  \ref{summaryfig2}-b), is an infinite-period bifurcation:  fixed any $\mu \!>\!\mu^*$, 
for $\gamma \!>\!\hat\gamma(\mu) \!\equiv\!1$ 
(grey area in fig.  \ref{summaryfig2}-b) the attracting pseudoperiodic path exists; for $\gamma \!\le\!1$ 
(white area in fig.  \ref{summaryfig2}-b)  it does not exists, and is replaced by a sequence of
infinite-period  saddle-node connections
for $\gamma \!=\!1$ (each leaving from a  $C_k$ and ending into $C_{k+1}$).
All these saddle-node connections again yield  unstable $\varphi$.

For $\gamma \!<\!\hat\gamma(\mu)$ (white area in fig.  \ref{summaryfig2}-b) neither 
the pseudoperiodic nor the heteroclininc paths
exist; the saddle connections leaving from $A_k$ end into $B_k$ or $B_{k+1}$. This implies 
that the corresponding $\varphi$ are unstable, because go to $\varphi^u$ either as $\xi\!\to\!\infty$, or as $\xi\!\to\!-\infty$.

The segment $\big(]0,\mu^*[,1\big)$ (the line separating the dark grey from the light grey region in  
fig.  \ref{summaryfig2}-b) is a saddle-node bifurcation of fixed points: fixed any $\mu \!>\!\mu^*$
any saddle-node $C_k$ transforms into  the pair node $(B_k,A_k)$  for  $\gamma \!<\!1$,
while it disappears for  $\gamma \!>\!1$.

\medskip
\noindent
We have thus partly proved the following theorem (assuming $\gamma\!\ge\! 0$
is no  loss of generality):

\begin{theorem} 
Assume $\alpha,\gamma\!\ge\!0$. Up to arbitrary translations of  $x,t$ and addition of multiples
of   $2\pi$, the relevant travelling-wave solutions of (\ref{equation}) are of the following types:

\begin{enumerate}

\item Static, uniform 
$\varphi^s(x,t;\gamma)\equiv \theta:=-\sin^{-1}\gamma$, for $\gamma<1$ 
(white region, dark grey region and red curve in fig.  \ref{summaryfig2}-b).

\item Kink $\hat\varphi^+$ or antikink $\hat\varphi^-$, where
$\hat\varphi^{\pm}(x,t;\gamma):=\hat g\left\{\xi^{\pm}\big[\hat\mu(\gamma)\big];\gamma\right\}\!-\!\pi$, for $\gamma\!<\!1$
(red curve in fig.  \ref{summaryfig2}-b).
$\hat\varphi^{\pm}$ respectively travel with phase velocity $v=\pm\check v\big(\hat\mu(\gamma)\big)$
and fulfill
\be
\lim\limits_{x\to -\infty}\!\hat\varphi^{\pm}(x,t;\gamma)=\theta,
\qquad\quad
\lim\limits_{x\to \infty}\!\hat\varphi^{\pm}(x,t;\gamma)=\theta\!\pm \!2\pi.              \label{asym1}
\ee

\item  Arrays of kinks $\check\varphi^+$ or antikinks $\check\varphi^-$, given by: 
$\check\varphi^{\pm}(x,t;\gamma,\mu):=\check g\big[\xi^{\pm}(\mu);\gamma,\mu\big]\!-\!\pi$,
for any  $\mu\!\in [0,\infty[$
and $\gamma\!\ge\! \hat\gamma(\mu)$ (dark and light grey regions in fig.  \ref{summaryfig2}-b);
$\check\varphi^{\pm}(x,t;\gamma,\infty)\!=\!\tilde g(\pm\! x\!-\!t;\gamma)\!-\!\pi$ 
if $\gamma\!>\!1$, $\mu\!=\!\infty$. $\check\varphi^{\pm}$
have resp. velocity $v\!=\!\pm\check v(\mu)$
and fulfill [with $\Xi$ as defined in  (\ref{linper})]
\be
\check\varphi^{\pm}(x\!+\!X\!,\!t;\gamma\!,\!\mu)=\check\varphi^{\pm}(x\!,\!t;\gamma\!,\!\mu)\pm 2\pi, \qquad 
X\!:=\!\left\{\!\!\ba{ll} \Xi\sqrt{1\!-\!v^2}\quad & \mbox{if }\mu\!<\!\infty\:\:\Leftrightarrow\:\:|v|\!<\!1,\\
 \frac{2\pi\alpha}{\sqrt{\gamma^2\!-\!1}}\quad & \mbox{if }\mu\!=\!\infty\:\:\Leftrightarrow\:\:|v|\!=\!1. \ea \right.
\label{circle}  \ee

\item Half-array of kinks
$\bar \varphi^+$  or antikinks $\bar\varphi^-$,  with
$\bar \varphi^{\pm}(x,t;\gamma\!,\!\mu)\!:=\!\bar g\big[\xi^{\pm}(\mu);\gamma\!,\!\mu\big]\!-\!\pi$,
only if $0\!<\!\gamma\!<\!1$ and for any $\mu\!\in ]0,\hat\mu(\gamma)[$ (dark grey region in fig.  \ref{summaryfig2}-b).
$\bar\varphi^{\pm}$ respectively have velocity $v=\pm\check v(\mu)$. They fulfill
\bea
&&\lim\limits_{x\to \mp\infty}\bar \varphi^{\pm}(x,t;\gamma\!,\!\mu)=\theta, \qquad\quad \lim\limits_{x\to \pm\infty}
[\bar\varphi^{\pm}(x,t;\gamma\!,\!\mu)-\check \varphi^{\pm}(x,t;\gamma\!,\!\mu)]=0^+, \label{lilla3}\\[8pt]
&& \lim\limits_{g\to \infty}[\bar z(g)\!-\!\check z(g)]=0^-\!,\qquad
\lim\limits_{\xi\to \infty}[\bar g'\!(\xi)-\check  g'\!(\xi)]=0^-\!,\qquad
 \lim\limits_{\xi\to \infty}[\bar g(\xi)\!-\!\check g(\xi)]=0^+\!,\qquad
\label{lilla2} 
\eea
for  suitable choices of $\check g,\check \varphi^{\pm}$ within their families
$[\check g], [\check \varphi^{\pm}]$ whose elements 
differ only by a $x$-translation. All limits are approached exponentially fast.

\end{enumerate}
\noindent
All of $\hat g, \check g,\bar g,\bar g\!-\! \check g$ are strictly increasing.
 To parameterize the solutions  of classes 3,4 one can adopt as an
independent variable alternative to  $\mu$  either  $z_M,\check I, |v|$ or $\Xi$.

\noindent
All other solutions $\varphi$  are manifestly unstable and/or have unbounded energy density $h$.

 \label{ThmClassif}
\end{theorem}

{\bf Remark \ref{PSG}.1} In Fig. \ref{summaryfig2}-a
we have plotted the qualitative behaviour of a kink, an array of kinks and
a half-array of kinks. The latter has no unperturbed analog. It
interpolates between the kink at one extreme and the
array of  kinks  at the other.
Therefore it cannot be approximated, nor can it even be figured out, by the
modulation Ansatz (\ref{modulation}).

\medskip
{\bf Remark \ref{PSG}.2} \
 $\check\varphi^{\pm}$ make
sense also as  solutions of (\ref{equation}) {\bf on
a circle of length $L=m X$}, for any $m\in\b{N}$. The integer $m$ parameterizes
different topological sectors: in the $m$-th sector the  pendula chain twists around
the circle $m$ times.

\medskip
{\bf Remark \ref{PSG}.3}
We emphasize that, in contrast with the unperturbed kink and array of
kinks, where $v$ was a free parameter of modulus less than 1,
$v$ is predicted as a function of $\gamma,\alpha$ in the  perturbed kink,
as a function of $\gamma,\alpha$ and one of the parameters $z_M,\check I,
\Xi$ in the  perturbed array and half-array of kinks.

\medskip
{\bf Rest of the proof:}
As $\check z(g_k^M)>0$, 
$\bar w=\bar z\!-\!\check z$ is negative-definite, and we find in the order
\be
\bar w:=\bar z-\check z\uparrow  0,\qquad\sqrt{2\bar
z}-\sqrt{2\check z}\uparrow 0, \qquad\frac 1{\sqrt{2\bar
z}}-\frac 1{\sqrt{2\check z}}\downarrow 0,      \label{limits1}
\ee
exponentially fast as $g\to\infty$. The first limit gives (\ref{lilla2})$_1$.
From (\ref{quadrature}) we obtain
$$
\bar\xi(g)=\!\int\limits^{ g}_{ g_0}\!\!\!
\frac{ds}{\sqrt{\!2\bar z(s)}}+\bar c, \qquad 
\check\xi(g)=\!\int\limits^{ g}_{ g_0}\!\!\!
\frac{ds}{\sqrt{\!2\check z(s)}}+\check c, \qquad 
\bar\xi(g)-\check\xi(g)=\!\int\limits^{ g}_{ g_0}\!\!\! ds\!\left[\frac{1}{\sqrt{\!2\bar
z(s)}}\!-\! \frac{1}{\sqrt{\!2\check z(s)}}\right]+(\bar c\!-\!\check c).
$$
where $\bar c,\check c$ are integration constants.
The last integrand is positive and goes exponentially to zero as $g\!\to\!\infty$,
hence the integral converges. Choosing \ $\bar c\!-\!\check c\!=\!\int^\infty_{ g_0}ds[]$  we find
$$
\bar\xi(g)=\check\xi(g)-\rho(g), \qquad\qquad \rho(g):=\int^{\infty}_{ g}ds
\left[\frac{1}{\sqrt{2\bar z(s)}}\!-\! \frac{1}{\sqrt{2\check z(s)}}\right]
$$
with $\rho(g)$
positive and exponentially vanishing. Applying  the inverse $\bar g(\xi)$
of $\bar\xi(g)$ to both sides we find
$$
g=\bar g\left[\check\xi(g)\!-\! \rho(g)\right]
=\bar g\left[\check\xi(g)\right]\!-\!  \bar g'(\tilde\xi)\rho(g).
$$
The second equality is based on Lagrange theorem, where $\tilde \xi$
is a suitable point in $]\check\xi(g)\!-\! \rho(g),\check\xi(g)[$.
Finally, setting $g=\check g(\xi)$ we find
$$
\check g(\xi)=\bar g\left(\xi\right)-\bar g'(\tilde
\xi)\rho\big[\check g(\xi)\big],
$$
where $\tilde \xi\in]\xi\!-\!\rho\big[\check g(\xi)\big],\xi[$.
The second term at the rhs exponentially vanishes as  $\xi\to \infty$ [since
$\rho\big(\check g(\xi)\big)$ does and $\bar g'$ is bounded],  proving
(\ref{lilla2})$_2$. By (\ref{lilla2})$_1$ now (\ref{limits1})$_2$ implies (\ref{lilla2})$_3$.

   Let $\tilde g(\eta):= \check g(\xi)$ with
$\eta:=\sqrt{1\!-\! v^2}\xi=\mbox{sign}(v)x\!-\!|v|t$.
$\tilde g_{\eta},\tilde g_{\eta\eta}$
are periodic. For  $\gamma\!>\!1$, replacing in (\ref{equation"})
and letting $|v| \uparrow 1$ we find that $\tilde g$ fufills (\ref{equa}). This
proves the limit
$ \lim_{\mu\to\infty}\check g\left(
\frac{\pm x\!-\!|\check v|t}{\sqrt{1\!-\!\check v^2}};\mu\right)=
\tilde g(\pm \!x\!-\!t)$, after noting that by (\ref{defmu})$_2$ $\mu\to \infty$ as  $|v| \uparrow 1$.

\medskip
Finally, we show that no other relevant solutions exist\footnote{In Ref. \cite{DanDeaFio05} this was 
shown only for $\gamma\!<\!1$. Actually the arguments used
there apply also for $\gamma\!\ge\!1$}. As already said in section \ref{rsGe},
if \ $\gamma\!=\!\alpha\!=\!\mu\!=\!0$ the other solutions with $|v|\!>\!1$ or ${\rm e}\!<\!1$
are unstable. If $\gamma\!>\!0$, this also applies to the solutions
arising from the cycles of (\ref{equation"}), if any.  
If $\gamma=1$ the paths connecting $C_k,C_{k\!+\!1}$  \cite{SanCon56} 
yield  manifestly unstable $\varphi$, in that they connect two unstable static solutions.	
If $\gamma\!>\!0$, $\alpha\!=\!\mu\!=\!0$,
the solutions $p(\xi)$ of (\ref{equation"}) which are unbounded in $g$
are unbounded also in $u$, by conservation of ${\rm e}$; if $\gamma\!>\!0$, $\alpha,\mu\!>\!0$,
this applies to all  unbounded solutions in $g$ except the pseudoperiodic $\check p(\xi)$,
 the saddle connections $\bar p_k(\xi)$\footnote{In fact, if $u(\xi)\!>\!0$ consider the $\check p(\xi)$
with argument $\xi$ shifted the right amount in order that it attracts $p(\xi)$ as $\xi\!\to\!\infty$. 
If $u(\xi)\!>\!\check u(\xi)$, by Property \ref{monoseq}
 $u(\xi)\!\to\!\infty$ as $\xi\!\to\!-\infty$; if $u(\xi)\!<\!\check u(\xi)$
then $p(\xi)$ either is a saddle connection $\bar p_k(\xi)$, or its $u(\xi)$ becomes negative for sufficiently
early 'times' $\xi$, and again by Property \ref{monoseq}
 $u(\xi)\!\to\!-\infty$ as $\xi\!\to\!-\infty$. The latter situation occurs also to the $p(\xi)$ with negative 
$u(\xi)$ for sufficiently early 'times' $\xi$ and ending on some $A_k,B_k$, or $C_k$.}.
The  $p(\xi)$ that are  unbounded in $u$ yield solutions $\varphi$ of (\ref{equation})
which have unbounded energy density $h$. Therefore, in all cases
they yield no other relevant  $\varphi$.
\hfill $\Box$

\smallskip
We finally determine   the ranges of the various parameters. Clearly, as $z_M\to \infty$ $\check z$ and
$\check I$ diverge, whereas $\check \mu,\Xi,\check v$ go to zero.
We now consider  the limit $z_M\to 0$.
If  $\gamma\!>\!1$, as $z_M\to 0$ one finds the following leading parts and limits
\be
\ba{ll}\check\mu\approx
\frac{\gamma-1}{\sqrt{2z_M}}\to\infty,\qquad\qquad  &\check I\approx
\frac{\sqrt{2z_M}2\pi\gamma}{\gamma-1}\to 0, \\[8pt] \Xi\sim
\frac 1{\sqrt{z_M}}\to\infty,\qquad\qquad &\check v\approx
\frac{\gamma-1}{\sqrt{2z_M\alpha^2+(\gamma-1)^2}}\to 0; \ea
\label{leading}
\ee
as $z_M$ spans $]0,\infty[$, the range of any of $\check I,\check\mu,\Xi$
is  $]0,\infty[$ and that of $\check v$ is $]0,1[$. In fact,  the Taylor formula of second order for
$\check z(g)$ around $g_k$ can be written without loss of generality in the form
\be
\check z(g;z_M;\gamma)=z_M+z_M\zeta_1(z_M;\gamma)(g-g_k)+(g-g_k)^2\rho(g)
\ee
with $\rho(g)$ bounded;
in order that, as $z_M\to 0$, $\check z$ keeps nonnegative  both in a left and a
right neighbourhood of $g_k$,  $\zeta_1(z_M;\gamma)$ has to approach
a {\it finite} limit. Replacing this Ansatz in (\ref{1stord}) we find at lowest
order in $(g-g_k)$
$$
z_M\zeta_1+\check\mu\sqrt{2z_M}+1-\gamma=0.
$$
As $z_M\to 0$ this implies (\ref{leading})
[by (\ref{EnBalance}), (\ref{linper})$_2$ and (\ref{arrvel})$_2$] .
Summarizing, as $z_M$ spans $]0,\infty[$ the
range of any of $\check I,\check\mu,\Xi$
is  $]0,\infty[$ and that of $\check v$ is $]0,1[$.

\noindent
If $\gamma\!\le\!1$,  by the monotonicity property $\check
\mu(\gamma,z_M)\le \hat\mu(\gamma)$, and by the continuity
we find \cite{Tri}
$$
\lim\limits_{z_M\to 0}\check \mu(\gamma,z_M)=\hat\mu(\gamma)<\infty.
$$
Hence if
$\gamma\le 1$ the range  of $\check\mu$ as $z_M$ spans $[0,\infty[$
is $]0,\hat\mu]$, the range  of $\check I$ is $]2\pi\gamma/\mu,\infty[$
the range  of $v$ is  $[0,\hat v[$.
The following bounds for $\hat \mu(\gamma)$  hold \cite{Tri,Hay53}
(see \cite{SanCon56} for a summary)
\be
\sqrt{\sqrt{3(1\!-\!\gamma^2)\!+\!1}\!-\!2\sqrt{1\!-\!\gamma^2}}\le\hat
\mu(\gamma)\le \sqrt{2\left(1\!-\!\sqrt{1\!-\!\gamma^2}\right)}.
\label{hatmubounds}
\ee

In \cite{Sco69} a theorem of linearized spectral (in)stability of travelling-wave solutions 
of sGe ($\alpha\!=\!\gamma\!=\!0$) was presented: the solutions with $|v|\!>\!1$ (fast solutions)
are always unstable; 
the solutions with $|v|\!<\!1$ (slow solutions) are spectrally stable if they are pseudoperiodic or of (anti)kink type.
The authors of  \cite{JonMarMilPla13} have detected and corrected an error in the proof.  
Note that this can lead at most to orbital stability for  $v\!\in]-1,1[$, because
no travelling-wave solutions of sGe can be stable or 
asimptotically stable in the strict sense.
In fact, the phase velocity $v$ for $\alpha\!=\!\mu\!=\!0$ is a free parameter; when replacing
 (\ref{redef'}) in a particular solution $g(\xi)$ of (\ref{equation"}) one obtains a 
family of solutions $\varphi(x,t)$ parametrized by $v$,
 like the (anti)kink ones (\ref{kink}). A small change in the initial conditions in general causes
a small change of $v$, which however leads to a  constantly growing deviation from
the initial solution, which will become larger and larger after a sufficiently long time.

One may expect that the situation improves in the perturbed case, because $v$ is 
determined by $\mu,\alpha,\gamma$.
In section 4 of \cite{DerDoeAvanVis03}  it is shown that the (anti)kink solution
of the perturbed equation (\ref{equation}) (with $\gamma\!<\!1$) is, up to a shift of the argument,
 asymptotically stable. A theorem  of asymptotic linearized stability for both the (anti)kink 
and the array of the (anti)kinks with $\gamma\!<\!1$ is proved in \cite{Mag80}  for $\gamma\!<\!1$; but only
w.r.t.  compact  variations of the initial conditions and in the sense of a pointwise convergence
to such solutions as $t\!\to\infty$.\footnote{Our $v,X(v)$ are resp. denoted as
$c,L(c)$ in \cite{Mag80}. Incidentally, the velocities 
$c_1\!<\!1$, $c_2\!>\!1$  of the slow and fast solitary waves considered there
are in fact the two solutions $v_1,v_2$ of (\ref{defmu}) seen as an equation in the unknown
$|v|$ when $\mu\!=\!\hat\mu(\gamma)$; as a consequence they fulfill the relation
$v_1^{-2}\!+\!v_2^{-2}\!=\!2$, not noted in \cite{Mag80}.}

\section{Method of successive approximations}
\label{successapprox}

Eq. (\ref{volterra}) can be reformulated as the fixed point equation
\be Az=z \label{fixedp}\ee for $z(g)$, where for $\epsilon>0$ the
operator $A=A(g_0,z_0;\mu,\gamma)$  is defined by \be\ba{l}
Aw(g):=\omega(g;g_0,z_0;\gamma)\!-\!
\int\limits_{g_0}^gds\, \phi\big(g,s,w(s)\big)\\[8pt]
\omega(g;g_0,z_0;\gamma):=z_0\!+\!U(g_0)\!-\!U(g)\qquad
\phi(g,s,\zeta):=\sqrt{2 \zeta}\mu \ea \ee on the space of
nonnegative smooth functions $w$ on $\mathbb{R}$ (the domain of $w$ can be
always trivially extended to $\mathbb{R}$). According to the method of
successive approximations, after a reasonable choice of a function
$z_{(0)}(g)$
 as an initial approximation for $z(g)$,  better and better
approximations should be provided by $z_{(n)}:=A^nz_{(0)}(g)$ as
$n\to\infty$. For this to make sense, at each step it is necessary
that $z_{(n)}$  belongs to the domain of $A$ (in the present case,
it must be nonnegative, otherwise the integrand function is
ill-defined) and that the sequence converges. {\it With the known
standard theorems}, this can be guaranteed a priori not in the whole
domain $G$ of the unknown $z$, but only in some smaller interval $J$
containing $g_0$. In general only the iterated application in
infinitely many adjacent intervals allows to extend a local solution
to a global one, what makes the procedure of little use for its
concrete determination.

Estimating the length of such a $J$ one
finds that it is not less than $2\pi$ only for sufficiently large
$z_0$. Actually, the determination of the solution in an interval of length
$2\pi$ would be enough for the complete determination both in the
case of a periodic solution $\check z$ (which is then extended
periodically) and of a separatrix $\hat z$ (in that case
$G=]g^M_{k\!-\!1},g^M_k[$, which has exactly length  $2\pi$). The periodicity condition
(\ref{periodicity}) is automatically fulfilled by each $z_{(n)}$ if we modify
the definition of $A$ adjusting the coefficient $\mu$ to $w$ as
follows:
\be
\tilde A \, w:=A\big(g_0,z_0;\tilde\mu(w),\gamma\big)\, w\qquad\qquad
\tilde\mu(w):=2\pi\gamma\left[\int\limits_{g_0}^{g_0\!+2\pi}ds\sqrt{2
w(s)}\right]^{-1}
\ee
Choosing $g_0=g^M_{k\!-\!1}$ for simplicity,
then $\tilde\mu(z_{(n)})$ will converge to $\check\mu(\gamma,z_0)$.
If instead we fix $\mu$ as an independent parameter, one will obtain
$z_0$ as $\lim_n z_{(n)}(g_0)$ \cite{Tri}. For the periodic solution
a sufficiently large $z_0$ amounts to a sufficiently small $\mu$; in
\cite{Tri} the following quantitative condition was found:
\be                                        \label{Tricomibound}
\eta_1>\epsilon_1, \qquad\mu<
\frac{(\sqrt{\eta_1}-\sqrt{\epsilon_1})^2}{2\pi\sqrt{2}}
\ee
where
$$
\epsilon_1:=\max|z_{(1)}\!-\!z_{(0)}|\equiv \Vert
z_{(1)}\!-\!z_{(0)}\Vert_{\infty},\qquad \eta_1:=\min |z_{(1)}|.
$$
So $\eta_1$ cannot be too small, in particular cannot vanish, what
excludes the cases of the periodic solutions $\check z$  having low energy
and of the heteroclinic path $\hat z$.

\subsection{The kink solution by the method of successive approximations}

The standard theorems fail for $\hat z$ because the sup norm has not
enough control to guarantee non-negativity of the approximations
$z_{(n)}$ everywhere in $G$, as well as the fulfillment of a Lipschitz
condition by the integrand $\phi$ and the behaviour (\ref{nearA_k})
near the extremes of $G$. In this section {\it we adopt a  clever,
nonstandard choice of the norm and show (Theorem \ref{born}) that a single
application of the method of successive approximations gives the kink
solution $\big(\hat \mu,\hat z(g)\big)$ in its whole domain
$G=]g^M_{k\!-\!1},g^M_k[$}.

Assume $\gamma<1$. Choose $g_0=g^M_{k\!-\!1}$, $z_0=0$ and let
$y:=g-g_0$. Then \be \omega(y)=\sqrt{1\!-\!\gamma^2}2\sin^2 \frac
y2+\gamma(y\!-\!\sin y)=\frac 12\sqrt{1\!-\!\gamma^2}y^2+O(y^3) \ee
and $\hat z$ fulfills  (\ref{fixedp}), where the operator $\tilde A$ has
taken the form \bea &&\tilde Az(y)\!\equiv\!\tilde
z(y)\!:=\!\sqrt{1\!-\!\gamma^2}2\sin^2\frac y2\!+\!\gamma(
y\!-\!\sin y)\!-\! \tilde\mu(z)\int\limits_0^ydy'\!\sqrt{2
z(y')},\nn &&\mbox{where
}\tilde\mu(z):=\frac{2\pi\gamma}{\int\limits_0^{2\pi}dy'\sqrt{2
z(y')}} \label{defA}\eea

By (\ref{nearA_k}) $\hat z(y)=O(y^2)$, $\hat
z(2\pi\!-\!y)=O\Big((2\pi\!-\!y)^2\Big)$. One easily checks that, more
generally, if $z$ has such a behaviour near $0,2\pi$, so has $\tilde Az$.
So it would be more natural to look for the solution from the very
beginning in a functional space whose elements have such a
behaviour. In $C^1([0,2\pi])$ introduce the norm \be \Vert z
\Vert=\sup_{y\in [0,2\pi]} \left|\frac{2z(y)}{ p^2(y)}\right|
\label{norm} \ee where the `weight' $p$ should vanish as $y$ and
$2\pi\!-\!y$ at $0,2\pi$ and will be specified later. Clearly \be \Vert
z \Vert \ge C\Vert z \Vert_{\infty}\equiv C\sup_{y\in
]0,2\pi[}|z(y)|\qquad\qquad C^{-1}:=\sup_{y\in
]0,2\pi[}\frac{p^2(y)}2.    \label{maggior} \ee The subspace \be
V:=\left\{z(y)\in C^{\infty}([0,2\pi]) \:\:|\:\: \Vert z
\Vert<\infty \right\} \ee is a complete metric space w.r.t. the
metric induced by the above norm. In fact, consider a Cauchy
sequence $\{z_n\}\subset V$ in the norm $\Vert\cdot\Vert$: by
(\ref{maggior}) it is Cauchy and therefore converges to a (uniformly
continuous) function $z(y)$ also in the norm
$\Vert\cdot\Vert_{\infty}$; moreover for any $\varepsilon>0$ there
exists $\bar r\in\b{N}$ such that $\forall r\ge \bar r$, $\forall
m\in\b{N}$ $$ \sup_{y\in [0,2\pi]} \left|\frac{z_r(y)-z_{r+m}(y)}{
p^2(y)}\right|< \frac{\varepsilon}2;
$$
Letting $m\to\infty$ we find
$$
\sup_{y\in [0,2\pi]} \left|\frac{z_r(y)-z(y)}{ p^2(y)}\right|< \varepsilon,
$$
showing that $z\!\in\! V$\footnote{If {\it ad absurdum} $\sup|z/p^2|=\infty$
then the lhs would certainly exceed $\varepsilon$.} and
$\{z_n\}\!\to\! z$ also w.r.t. the topology induced by the above norm.

Let $a,b\in\mathbb{R}$ with $b>a>0$. The subset \be
Z_{a,b,p}:=\left\{z(y)\in V\:\:|\:\:  a^2\le \frac{2z(y)}{
p^2(y)}\le b^2 \right\} \ee is clearly closed w.r.t. the metric
induced by the above norm. We shall look for $(\hat z,\hat\mu)$
within a suitable $Z_{a,b,p}$. First we look for $a,b$ such that
(\ref{defA}) defines an operator $\tilde A:Z_{a,b,p}\to Z_{a,b,p}$. Up to a
factor, we choose $p^2(y)$ as the $\gamma=0$ (i.e. unperturbed)
kink solution $\hat z_0(y)$, more precisely $p(y):=\sin\frac y2$.
Then
$$
P(y):=\int\limits_0^ydy'p(y')=2(1-\cos\frac y2)
=\int\limits_y^{2\pi}dy'p(y'),
$$
and, since $1-\sqrt{1-w}\ge w/2$  we find (setting $w=\sin^2\frac y2$)
$$
p^2(y)\le P(y)\le 2\left(1\!-\!\cos\frac y2\right)\left(1\!+\!\cos\frac
y2\right)=
2p^2(y).
$$
Thus for any $z\in Z_{a,b,p}$
we find
\bea
&& aP(y)\le\int\limits_0^ydy'\sqrt{2 z(y')}=
\int\limits_0^ydy'\frac{\sqrt{2 z(y')}}{p(y')}p(y')\le b
P(y) \nn
&&4a=
aP(2\pi)\le\frac{2\pi\gamma}{\tilde\mu}=\int\limits_0^{2\pi}dy'\sqrt{2 z(y')}=
\le b P(2\pi)=4b\nonumber \eea
implying
the inequalities $\gamma\pi/2b\le\tilde\mu\le
\gamma\pi/2a$ and
\be
 \gamma \frac{\pi a}{2b}p^2(y) \le  \tilde\mu\int\limits_0^ydy'\sqrt{2
z(y')}\le   \gamma \frac{\pi b}{a}p^2(y).          \label{magg1}
\ee
Similarly,
\be
 \gamma \frac{\pi a}{2b}p^2(y) \le  \tilde\mu\int\limits_y^{2\pi}dy'\sqrt{2
z(y')}\le   \gamma \frac {\pi b}{a}p^2(y).           \label{magg1'}
\ee

\begin{lemma} For all $y\ge 0$
$$
1-\cos y\ge 0, \qquad y-\sin y\ge 0, \qquad
\frac{y^2}2-1+\cos y\ge 0, \qquad\frac{y^3}6-y+\sin y\ge 0.
$$
\end{lemma}
Proof: The first equality is obvious; the other ones follow by integrations over $[0,y]$. QED.

As a consequence,
 for $y\in [0,\pi]$
\be
0\le y\!-\!\sin y\le \frac{y^3}6=\frac 16 p^2(y)\left[\frac{y}{\sin\frac y2}
\right]^2\! y\le p^2(y)\frac{\pi^3}6.                     \label{maggior2}
\ee
Collecting the results,  on one hand assuming $1\ge a/b\ge 1/2$
we find
\be
\tilde z(y) \ge p^2(y)\left[2\sqrt{1\!-\!\gamma^2}
-\gamma\pi \frac ba\right] \ge
 p^2(y) 2\left[\sqrt{1\!-\!\gamma^2}
-\gamma\pi \right]
\ee
for all  $y\in [0,2\pi]$; on the other hand, for $y\in [0,\pi]$ we find
\be
\tilde z(y)\le p^2(y)2\left[\sqrt{1\!-\!\gamma^2}+\gamma\frac {\pi^3}{12}\right].
\ee
This provides bounds for $y\in [0,\pi]$. To find bounds for $y\in [\pi,2\pi]$
set $v=(2\pi\!-\!y)$ and note that from (\ref{defA}) it follows
\bea
\tilde z(y) &=& \sqrt{1\!-\!\gamma^2}\,2\sin^2\frac y2\!-\!\gamma(
v\!-\!\sin  v)\!+\!2\pi\gamma-\!
\tilde\mu\!\left[\int\limits_0^{2\pi}\!dy\sqrt{2
z}\!-\!\int\limits_y^{2\pi}\!dy'\sqrt{2 z(y')}\right]
\nn
&=&\sqrt{1\!-\!\gamma^2}\,2\sin^2\frac y2\!-\!\gamma( v\!-\!\sin  v)\!+\!
\tilde\mu\int\limits_y^{2\pi}dy'\sqrt{2 z(y')}, \nonumber
\eea
We use (\ref{magg1'}) to bound the third term at the rhs;
as $v\in[0,\pi]$, to bound the second term we can use (\ref{maggior2})
with $y$ replaced by $v$, but keeping $p^2(y)=p^2(v)$ at the rhs of
the latter.
Collecting the results we thus find  for $y\in [\pi,2\pi]$
\be
p^2(y)2\left[\sqrt{1\!-\!\gamma^2}-\gamma\frac {\pi^3}{12}\right]\le\tilde
z(y)\le p^2(y) 2\left[\sqrt{1\!-\!\gamma^2}+\gamma\pi\right].
\ee
Hence $a^2p^2\le 2\tilde z\le b^2p^2$, so that $\tilde z\in  Z_{a,b,p}$,  if we define
\be
a^2:=4\left[\sqrt{1\!-\!\gamma^2}-\gamma\pi\right],\qquad\qquad
b^2:=4\left[\sqrt{1\!-\!\gamma^2}+\gamma\pi\right].
\ee
In order that $1/2\le a/b$ it must be
$$
\frac 14 \le\frac {a^2}{b^2}=\frac
{\sqrt{1\!-\!\gamma^2}-\gamma\pi}{\sqrt{1\!-\!\gamma^2}+\gamma \pi}
$$
what gives, after some computation,
\be
\gamma\le\left[1+\frac{25\pi^2}9\right]^{-\frac 12}
\approx .187
\ee
We conclude that in this $\gamma$-range with the above choice of $a,b$
$\tilde AZ_{a,b,p}\subset Z_{a,b,p}$, as required.

\medskip
Let us determine the constraints on $a,b$ following from the condition that $\tilde A$
be a contraction. First, we immediately find
$$
2|z_1(y)-z_2(y)|=p^2(y)\frac{2|z_1(y)-z_2(y)|}{p^2(y)}\le p^2(y)\Vert
z_1-z_2\Vert $$
Note that for any $\alpha>0$,
$|\sqrt{u_1}-\sqrt{u_2}|\le |u_1-u_2|/(2\alpha)$ if $u_1,u_2\in
[\alpha^2,\infty[$. Hence
\bea
|\sqrt{2z_1(y)}-\sqrt{2z_2(y)}|&=&p(y)\left|\sqrt{\frac{2z_1(y)}{p^2(y)}}-
\sqrt{\frac{2z_2(y)}{p^2(y)}}\right|\le
\frac{p(y)}{2a}\frac{2|z_1(y)-z_2(y)|}{p^2(y)}\nn
&\le& \frac{p(y)}{2a}\Vert z_1-z_2\Vert
\eea
\bea
|\tilde \mu_1-\tilde \mu_2| &=& \tilde \mu_1\tilde \mu_2|\tilde
\mu_1^{-1}-\tilde \mu_2^{-1}|\le\frac{\pi\gamma}{8a^2}
\left|\int\limits_0^{2\pi}dy(\sqrt{2 z_1(y)}-\sqrt{2
z_2(y)}\right|\nn
&\le&\frac{\pi\gamma}{8a^2}\int\limits_0^{2\pi}dy\left|\sqrt{2
z_1(y)}-\sqrt{2 z_2(y)}\right|\le\frac{\pi\gamma}{16a^3}\Vert z_1-z_2\Vert
\int\limits_0^{2\pi}dyp(y) \nn
&=&\frac{\pi\gamma}{4a^3}\Vert z_1-z_2\Vert    \label{mubounds}
\eea

$$
\tilde z_2-\tilde z_1=\int\limits_0^ydy'
\left[(\tilde \mu_1-\tilde \mu_2)\sqrt{2 z_1(y')}
+\tilde \mu_2(\sqrt{2z_1(y')}-\sqrt{2z_2(y')})\right]
$$
whence
\bea
|\tilde z_1(y)\!-\!\tilde z_2(y)| &\le& |\tilde \mu_1\!-\!\tilde
\mu_2|  \int\limits_0^ydy'
p(y')\sqrt{\frac{2 z_1(y')}{p^2(y')}}
+\tilde \mu_2\int\limits_0^ydy'\left|\sqrt{2z_1(y')}\!-\!\sqrt{2z_2(y')}\right|
\nn &\le& \frac{\pi b\gamma}{4a^3}\Vert z_1-z_2\Vert P(y)
+\frac{\pi\gamma}{4a^2}\Vert z_1-z_2\Vert P(y)
\nn &\le& \left(1+\frac ba\right)\frac{\pi\gamma}{4a^2}\Vert z_1-z_2\Vert P(y)
\le \left(1+\frac
ba\right)\frac{\pi\gamma}{2a^2}\Vert z_1-z_2\Vert p^2(y), \nonumber
\eea
implying
\be
\Vert\tilde z_1(y)-\tilde z_2(y)\Vert\le \left(1+\frac
ba\right)\frac{\pi\gamma}{a^2}\Vert z_1-z_2\Vert.
\ee
Thus, $\tilde A$ is a contraction if
\be
\lambda:=(1+b/a)\pi\gamma/a^2<1,           \label{lambdadef}
\ee
that is,
$$
\gamma<\frac{a^2}{\pi\left(1+\frac ba\right)}\le
\frac{a^2}{3\pi}= \frac{4}{3\pi}[\sqrt{1\!-\!\gamma^2}-\gamma\pi],
$$
namely if
\be
\gamma<\left[1+\left(\frac {7\pi}4\right)^2\right]^{-\frac 12}\approx .179
                                                    \label{gammarange}
\ee Summing up, under this condition $\tilde A$ is a contraction of
$Z_{a,b,p}$ into itself. Since $z_{(0)}(y):=2p^2(y)=2\sin^2\frac y2$
belongs to $Z_{a,b,p}$, applying the Banach fixed point theorem we
find

\begin{theorem}   Let $z_{(0)}(y)\!:=\!2\sin^2\frac y2$,
$z_{(n)}\!:=\!\tilde A^nz_{(0)}$,
$\mu_n\!:=\!\tilde\mu\big(z_{(n\!-\!1)}\big)$, with
$\tilde A,\tilde \mu$ defined as in (\ref{defA}). The sequences $\{z_{(n)}\}_{n\in\b{N}}$,
$\{\mu_n\}_{n\in\b{N}}$ converge respectively to the
kink solution $\hat z$ [in the norm (\ref{norm})]
and to the corresponding $\hat\mu(\gamma)$, for $\gamma$
at least in the range (\ref{gammarange}). With $\lambda$ defined as
in (\ref{lambdadef}), the errors of the $n$-th
approximation are bound by
\be \left\Vert z_{(n)}-\hat z\right\Vert
\le  \frac{\lambda^n}{1\!-\!\lambda}\left\Vert  z_{(1)}\!-\!
z_{(0)}\right\Vert,\qquad\qquad
|\mu_n-\hat\mu| \le \frac{\pi\gamma}{32}\!\left[\sqrt{1\!-\!\gamma^2}\!-\!\gamma\pi\right]^{-\frac 32}
\!\frac{\lambda^n}{1\!-\!\lambda}\left\Vert z_{(1)}\!-\! z_{(0)}\right\Vert\qquad
\ee
\label{born}
\end{theorem}
[To complete the proof we need just to note that, by (\ref{mubounds}), the convergence of
$z_{(n)}$ implies the convergence of $\mu_n$
and estimate the second error through standard arguments].

\medskip
{\bf Remark \ref{successapprox}.1} More refined computations of upper and
lower bounds, with the present $\gamma$-independent weight
$p^2(y)=\sin^2\frac y2$, would show a $\gamma$-range of convergence
of the above sequences slightly larger than (\ref{gammarange}).
By choosing a suitable $\gamma$-dependent weight
$p^2(y)$, e.g. $p^2(y)=z_{(1)}(y)/2$, one could show
that this range is actually significantly larger.
This will be elaborated elsewhere.

\bigskip

We explicitly work out  the first approximation. We find:
\bea
&& z_{(1)}(y)=\sqrt{1\!-\!\gamma^2}2\sin^2\frac y2\!+\!\gamma
\left[\pi\left(\cos\frac y2\!-\!1\right)\!+\!y\!-\!\sin y\right]\\[8pt]
&&\mu_1=\frac 14 {\pi\gamma} \\[8pt]
&& {\rm e}_{(1)}(y)=\gamma\pi\left(\cos\frac y2\!-\!1\right)\!+\!\mbox{const}
\\[8pt]
&& v_{(1)}(\gamma,\alpha)=\frac 1{\sqrt{1+(4\alpha/\pi\gamma)^2}}=\frac
{\pi\gamma}{4\alpha}+O(\gamma^2).                                 \label{v_1}
\eea
The results are in good agreement with the plot in Fig. \ref{summaryfig1} (right). Note that the result
(\ref{v_1}) coincides with (\ref{v_atinf}), as announced.
In a similar way one can determine iteratively solutions of type 3 ($\mu,\check
z$) even with low $z_M$ [i.e. not fulfilling the bound (\ref{Tricomibound})].


\subsection*{Acknowledgments}

We are grateful to C. Nappi for information on the present
state-of-the-art of research on the Josephson effect
and for useful discussions. It is also a pleasure to thank
Prof. A. D'Anna and P.
Renno for their encouragement and stimulating observations.
This research was partially supported
by UniNA and Compagnia di San Paolo under the grant "STAR Program 2013".

\section*{Appendix}
\label{Appsing}

{\bf Proof of Prop. \ref{mono2}.}
Let $0\!\le\!z_{0,2}\!<\!z_{0,1}$,  $z_j(g):=z(g;g_0,z_{0,j};\mu,\gamma)$
($j=1,2$) be
 the corresponding  solutions of (\ref{1stord})
and $G_j$ the corresponding intervals giving their (maximal)
domains. By continuity the inequality \be z_1-z_2>0
\label{ineq} \ee will hold in a neighbourhood of $g_0$ within
$G_1\cap G_2$.  In fact, it will hold for all $g\in G_1\cap G_2$. If
{\it ad absurdum} this were not the case, denote by $\bar g\in
G_1\cap G_2$ the least $g\!>\!g_0$ (resp. largest $g\!<\!g_0$) where
$z_1\!-\!z_2$ vanishes: $z_1(\bar g)\!-\!z_2(\bar g)\!=\!0$; then
the problem (\ref{1stord}) with initial (resp. final) condition
$z(\bar g)\!=\!z_1(\bar  g)\!\equiv\! z_2(\bar g)$ would admit the
two different solutions $z_1, z_2$, against the existence and
uniqueness theorem.  As for the monotonicity of $g_{\pm}$,
by the same theorem $z_1(g_{2\pm})\!>\!z_2(g_{2\pm})\!=\!0$ implies
$g_{1+}\!>\!g_{2+}$ if $g_{2+}\!<\!\infty$, otherwise $g_{1+}\!=\!g_{2+}\!=\!\infty$,
and $g_{1-}\!<\!g_{2-}$ if $g_{2-}\!>\!-\infty$, otherwise
$g_{1-}\!=\!g_{2-}\!=\!-\infty$.

\medskip
{\bf Proof of Prop. \ref{mono1}.}
Let $\mu_1\!\le\!\mu_2$, $\gamma_1\epsilon\!\ge\!\gamma_2\epsilon$,
with one of the two inequalities being strict; for $j=1,2$ let
$u_j(g):=u(g;g_0,u_0;\mu_j,\gamma)$  be the corresponding solutions
of (\ref{1stord}) with the same condition $u_j(g_0) =u_0$,
and  $G_j$ the intervals giving their (maximal) domains. We find
$$
u_{2g}=-\mu_2+\frac{\gamma_2-\sin g}{u_2}<-\mu_1+\frac{\gamma_1-\sin g}{u_2}.
$$
By the comparison principle\footnote{Here we recall the latter in
the restricted version: if $f$ fulfills conditions ensuring that
the differential problem $\tilde u'\!=\!f(x,\tilde u)$, $\tilde
u(x_0)=u(x_0)$, has a unique solution $\tilde u$, and $u'\!<
\!f(x,u)$ for all $x$, then it is $u(x)<\tilde u(x)$ for all
$x\!>\!x_0$ and $u(x)>\tilde u(x)$ for all $x\!<\!x_0$.} (see e.g.
\cite{Yos66}) it follows, as claimed, \be u_1(g)>u_2(g)\quad
g\in]g_0,g_+[,\qquad\qquad u_1(g)<u_2(g)\quad g\in]g_-,g_0[. \ee If
$\epsilon>0$, this implies: $\lim_{g\downarrow g_{2+}}u_1(g)\ge
\lim_{g\downarrow g_{2+}}u_2(g)=0$ and therefore $g_{1+}\ge
 g_{2+}$ (the
inequalities being strict as long as $g_{2+}<\infty$);
$\lim_{g\uparrow g_{1-}}u_2\ge 0$ and therefore $g_{1,-}\ge
 g_{2-}$
(the inequalities being strict as long as $g_{1-}>-\infty$).
Moreover, let $g_j(\xi)=g(\xi;g_0,u_0;\mu_j,\gamma_j)$  be the
corresponding two solutions of (\ref{1storder'}), i.e. the solutions
of (\ref{equation"}). We find
$$
g_2'(\xi)=u_2\big(g_2(\xi)\big)
\left\{\ba{l}
<u_1\big(g_2(\xi)\big),\qquad \forall \xi>\xi_0,\\[8pt]
>u_1\big(g_2(\xi)\big),\qquad \forall \xi<\xi_0,\ea\right.
$$
while $g_2(\xi_0)=g_0=g_1(\xi_0)$. By the comparison principle
this implies as claimed
 $g_2(\xi)<g_1(\xi)$  for all $\xi\in X_1\cap X_2$.
Similarly one argues if  $\epsilon<0$.

\medskip
{\bf Proof of Prop. \ref{monoseq}.}
Consider the Cauchy problem (\ref{1stord}) in
subsequent intervals   $]g_k,g_{k\!+\!1}[ \subset G$. Since  the equation
is invariant under $g\to g\!+\!2\pi$,  by
Prop. \ref{mono2}  if $z(g_1)$ is
respectively larger, equal, smaller than $z(g_0)$   then so are
$z(g_{k\!+\!1}),I_{k\!+\!1}$ in comparison  with $z(g_k),I_k$
respectively, for all $k\in K$;   in other words, the sequences
$\{z(g_k)\}$, $\{I_k\}$ are either constant, or strictly monotonic.
Eq. (\ref{bound}) follows from (\ref{volterra}) applied in
$]g_k,g_{k\!+\!1}[$.

If  $\epsilon\!=\!-$, then rhs(\ref{bound})$>2\pi\gamma>0$ for any $k$, so
that the sequences are strictly increasing and  diverging as $k\to\infty$,
whereas $K$ must have a lower bound, otherwise $z(g_k)$ would become
negative for sufficiently low $k$.

If  $\epsilon\!=\!+$, then the two terms at the rhs(\ref{bound}) have opposite
sign and can compensate each other. If  the sequences are strictly increasing, the
sides of (\ref{bound}) are positive for all $k$ and
$I_k<2\pi\gamma/\mu$.
Applying (\ref{volterra}) to the interval
$[g_k,g_k\!+\!\Delta g]$ for any $\Delta g\le 2\pi$ we find
$$
z(g_k\!\!+\!\!\Delta g)-z(g_k)=U(g_k)\!-\!U(g_k\!+\!\Delta
g)\!-\!\mu\int\limits_{g_k}^{g_k\!+\!\Delta g}ds\sqrt{2z(s)}.
$$
But $|U(g_k)\!-\!U(g_k\!+\!\Delta g)|$ is upper bounded, e.g. by
$2\!+\!2\pi\gamma$, whence
\be
|z(g_k\!\!+\!\!\Delta g)-z(g_k)|\le\!2\!+\!2\pi\gamma+\mu I_k< 2\!+\!4\pi\gamma.
\label{boundedness}
\ee
If {\it ad absurdum} $z(g_k)$
diverged as $k\to\infty$, then also  $z(g_k\!\!+\!\!\Delta g)$ and in turn
$I_k$ [by (\ref{defint})] would diverge, in contrast with $I_k<2\pi\gamma/\mu$;
so it must converge. Moreover, as before, $K$ must have a lower bound.
On the other hand, rewriting (\ref{bound}) in the form
$z(g_{k\!-\!1})-z(g_k)=\mu I_{k\!-\!1}-2\pi\gamma$, we see that
if the sequences $\{z(g_k)\},\{I_k\}$ are strictly decreasing, the sides are positive
for all $k$ and larger than $ \mu I_0-2\pi\gamma>0$ for all negative $k$;
this implies that they diverge as $k\to -\infty$, and again by (\ref{boundedness})
so do  $z(g),I(z,g)$. Whereas they must either converge as $k\to \infty$, or
$K$ must have an upper bound.

\end{document}